\documentclass[sigconf]{acmart}

\usepackage[T1]{fontenc}
\usepackage{graphicx}
\usepackage{hyperref}
\usepackage{amsfonts} 
\usepackage{amsmath}
\usepackage{multirow}
\usepackage{listings}
\usepackage{float}
\usepackage{xspace}
\usepackage{mathtools}
\usepackage{textcomp}
\usepackage{natbib}
\usepackage[capitalise,noabbrev]{cleveref}
\usepackage{enumitem}
\usepackage{listings}
\usepackage{pifont}
\usepackage{subcaption}

\AtBeginDocument{\DeclareCaptionSubType{lstlisting}}

\keywords{Constant time code, cryptographic implementations, compilers}

\ccsdesc[500]{Security and privacy~Systems security}
\ccsdesc[500]{Security and privacy~Software and application security}

\title{Breaking~Bad: How Compilers Break Constant-Time~Implementations}

\author{Moritz Schneider}
\orcid{0000-0002-8069-9848}
\email{moritz.schneider@inf.ethz.ch}
\affiliation{%
    \institution{ETH Zurich}
    \city{Zurich}
    \country{Switzerland}
}
\author{Daniele Lain}
\orcid{0000-0001-6101-7306}
\email{daniele.lain@inf.ethz.ch}
\affiliation{%
    \institution{ETH Zurich}
    \city{Zurich}
    \country{Switzerland}
}
\author{Ivan Puddu}
\orcid{0000-0003-2198-2405}
\email{ivan.puddu@inf.ethz.ch}
\affiliation{%
    \institution{ETH Zurich}
    \city{Zurich}
    \country{Switzerland}
}
\author{Nicolas Dutly}
\orcid{0009-0002-0829-6320}
\email{ndutly@student.ethz.ch}
\affiliation{%
    \institution{ETH Zurich}
    \city{Zurich}
    \country{Switzerland}
}
\author{Srdjan Čapkun}
\orcid{0000-0001-5727-3673}
\email{srdjan.capkun@inf.ethz.ch}
\affiliation{%
    \institution{ETH Zurich}
    \city{Zurich}
    \country{Switzerland}
}

\copyrightyear{2025}
\acmYear{2025}
\setcopyright{cc}
\setcctype{by}
\acmConference[ASIA CCS '25]{ACM Asia Conference on Computer and Communications Security}{August 25--29, 2025}{Hanoi, Vietnam}
\acmBooktitle{ACM Asia Conference on Computer and Communications Security (ASIA CCS '25), August 25--29, 2025, Hanoi, Vietnam}\acmDOI{10.1145/3708821.3733909}
\acmISBN{979-8-4007-1410-8/2025/08}

\newif\ifsubmission{}
\submissiontrue

\newif\ifprintheader
\printheaderfalse

\newif\ifdraft{}

\drafttrue{}

\newcommand{\nrexperiments}{44,604\xspace}

\newcommand{\yes}{\ding{52}}
\newcommand{\no}{\textcolor{gray}{\ding{55}}}
\newcommand{\blind}{\ding{83}\xspace}
\newcommand{\asm}{\ding{45}\xspace}

\ifdraft

\printheadertrue

\newcommand{\todo}[1]{\textcolor{red}{TODO: #1}}

\newcommand{\ivan}[1]{\textbf{\emph{ #1 \colorbox{magenta}{[Ivan]}}}}
\newcommand{\moritz}[1]{\textbf{\emph{ #1 \colorbox{yellow}{[Moritz]}}}}

\newcommand{\srdjan}[1]{\textbf{\emph{ #1 \colorbox{blue}{\textcolor{white}{[Srdjan]}}}}}

\else

\newcommand{\todo}[1]{}

\newcommand{\ivan}[1]{}
\newcommand{\moritz}[1]{}
\newcommand{\srdjan}[1]{}

\fi

\definecolor{codegreen}{HTML}{12711c}
\definecolor{codegray}{HTML}{949494}
\definecolor{codeblue}{HTML}{001c7f}
\definecolor{codeyellow}{HTML}{b1400d}

\definecolor{backcolour}{rgb}{0.96,0.96,0.94}

\lstdefinestyle{mystyle}{
    backgroundcolor=\color{backcolour},   
    commentstyle=\color{codegreen},
    keywordstyle=\color{codeyellow},
    keywordstyle={[2]\color{codeblue}},
    numberstyle=\tiny\color{codegray},
    stringstyle=\color{codeblue},
    basicstyle=\ttfamily\footnotesize,
    breakatwhitespace=false,         
    breaklines=true,                 
    captionpos=b,                    
    keepspaces=true,                 
    numbers=left,                    
    numbersep=5pt,                  
    showspaces=false,                
    showstringspaces=false,
    showtabs=false,                  
    tabsize=2,
    frame=ltb,
    framerule=0pt,
}

\lstdefinelanguage{asm}
    {morekeywords={mov,add,cmp,lw,sw,ld,sd,move,ldr,str,bx,and,srli,beqz,li,slli,xor,ret,test,lea,cmove,andi,movne},
    sensitive=false,
    morecomment=[l]{//},
    morecomment=[s]{/*}{*/},
}

\lstdefinelanguage{myc}
    {morekeywords={const,return,this,for,if,else},
    morekeywords={[2]unsigned,char,int,uint32_t,uint64_t,uint128_t,void},
    sensitive=false,
    morecomment=[l]{//},
    morecomment=[s]{/*}{*/},
}

\colorlet{tablerowcolor}{blue!3}

\begin{document}

\begin{abstract}
The implementations of most hardened cryptographic libraries use defensive programming techniques for side-channel resistance. These techniques are usually specified as guidelines to developers on specific code patterns to use or avoid. Examples include performing arithmetic operations to choose between two variables instead of executing a secret-dependent branch. However, such techniques are only meaningful if they persist across compilation. In this paper, we investigate how optimizations used by modern compilers break the protections introduced by defensive programming techniques. Specifically, how compilers break high-level constant-time implementations used to mitigate timing side-channel attacks. We run a large-scale experiment to see if such compiler-induced issues manifest in state-of-the-art cryptographic libraries. We develop a tool that can profile virtually any architecture, and we use it to run trace-based dynamic analysis on \nrexperiments different targets. Particularly, we focus on the most widely deployed cryptographic libraries, which aim to provide side-channel resistance. We are able to evaluate whether their claims hold across various CPU architectures, including x86-64, x86-i386, armv7, aarch64, RISC-V, and MIPS-32. 

Our large-scale study reveals that several compiler-induced secret-dependent operations occur within some of the most highly regarded hardened cryptographic libraries -- even when the high-level source code was formally verified to be free of side channels. To the best of our knowledge, such findings represent the first time these issues have been systematically observed in the wild and provide concrete data confirming previous speculations about defensive programming techniques' limitations. One of the key takeaways of this paper is that the state-of-the-art defensive programming techniques employed for side-channel resistance are still inadequate, incomplete, and bound to fail when paired with the optimizations that compilers continuously introduce.
\end{abstract}

\maketitle

\section{Introduction}

Since the discovery of timing attacks~\cite{kocher1996timing}, side-channel vulnerabilities have been one of the major concerns for developers of security-critical code and libraries~\cite{jancar2022they}. %
Particular attention and expert knowledge are devoted to avoiding side-channel issues in security-critical libraries. 
Three main hardening techniques approaches are generally followed: i) manual assembly hardening, ii) using special compilers that provide constant time guarantees, and iii) hardening the source code. However, all of these approaches suffer from practical limitations.

The first option is vetting hand-written assembly either by a developer~\cite{openssl} or with automated tools~\cite{bond2017vale}. However, this limits code portability and requires manual effort for each supported (micro-)architecture---in practice, only the most popular architectures are analyzed in this manner (e.g., x86~\cite{bond2017vale}). However, security-critical libraries are expected to be deployed everywhere, leaving less vetted architectures as second-class citizens in terms of security and hence potentially more susceptible to attacks. 
A solution to this problem is to compile the portable source code of security-critical libraries with special compilers that automatically remove side channels~\cite{borrello2021constantine,rane2015raccoon}. However, these compilers
suffer from a set of shortcomings: support for processor architectures is poor, they might require expert knowledge (e.g., to annotate the code), and they struggle to provide support for modern features of the processor or the employed source code languages. As a consequence, this approach is rarely used in practice, e.g., the binaries of security-critical libraries provided by Linux packaging repositories are compiled with commodity compilers such as \texttt{GCC} and \texttt{LLVM}. %
The third and final approach relies on hardening the higher-level source code using defensive programming techniques such as constant-time programming and then compiling the hardened code using commodity compilers. The advantage of this approach is that commodity compilers, as opposed to special compilers, offer support for many architectures, are maintained and get improvements by hundreds of developers, and support the latest CPU features.
Some projects even go as far as formally verifying the hardened higher-level code~\cite{zinzindohoue2017hacl}. 
However, commodity compilers might apply transformations or optimizations that re-introduce side-channel vulnerabilities---an issue that surveyed developers of security-critical libraries are aware and afraid of~\cite{jancar2022they}, and that has been previously observed in small manually crafted examples~\cite{simon2018you,daniel2020binsec}.

As a practical compromise between the three techniques mentioned so far, modern security-critical libraries rely on defensive programming techniques on high-level source code and the occasional spot check in their resulting binaries\cite{jancar2022they}. While there exist multiple recommendations on how to write hardened high-level source code~\cite{aumasson2012cryptographycodingstandard,aumasson2019cryptocoding}, and these have been applied to various projects~\cite{bearssl,botan,libsodium,mbedtls}, it remains unclear whether they are effective at preventing  \emph{compiler-induced} side-channels. Furthermore, most of these coding recommendations were proposed before 2010, and thus, they may no longer apply to modern compilers (and/or need updates for novel optimizations). Moreover, new processor architectures differ significantly, and compilers may leverage different architecture-specific optimization techniques. Hitherto, it has not been studied how robust these recommendations are w.r.t. modern compilers and build environments.

In this paper, we study arguably some of the most hardened projects: cryptographic libraries, as they need to handle long-term secrets. Cryptographic libraries are usually developed by experts with a particular sensibility towards security and defensive programming. Many such libraries consider side-channel attacks in their threat models and aim to be side-channel free~\cite{bearssl,libsodium,zinzindohoue2017hacl,boringssl,botan}, and their developers often diligently follow hardening practices. These implementations are often called constant-time to indicate that they behave exactly the same for different secret inputs. Some libraries even provide formal guarantees for the absence of side channels~\cite{zinzindohoue2017hacl}, while others use various tools and manual analysis to eliminate side channels~\cite{libsodium}. 

To understand if such constant-time development practices survive compilation, we ran a large-scale experiment to see if compiler-induced side-channel vulnerabilities exist in constant-time implementations used in the wild. We further aimed to understand how often they appear, which kind, and the reasons thereof.
To answer this, we developed a scalable pipeline that allows us to analyze different compiler options for any specified target algorithm.
We used our approach to perform a systematic study of the final binary of eight widely used cryptographic libraries on six different processor architectures, compiled with several versions of two common off-the-shelf compilers (GCC and LLVM) and with different optimization levels. This results in $6608$ distinct binaries. Each binary contains multiple cryptographic primitives (e.g., AES, ECDSA. etc.) resulting in the total of \nrexperiments targets to test (i.e., experiments)\footnote{Each binary contains multiple cryptographic primitives. For instance, one of the $6608$ binaries is OpenSSL compiled with LLVM 8 using \texttt{-O2} for x86-64. The compiled binary contains implementations for AES-GCM, HMAC-SHA1 etc. One experiment consists of testing AES-GCM from this binary. Another experiment consists of testing HMAC-SHA1 from the same binary, etc. }.
In each experiment, we tested a different binary and a different cryptographic primitive for their side-channel susceptibility. 
We used trace-based dynamic analysis~\cite{he2020ct,weiser2018data} to detect secret-dependent control flow and memory accesses, the two most prominent root causes of side-channel vulnerabilities. 

We discover many novel compiler-induced secret-dependent operations in state-of-the-art cryptographic libraries. These findings show that previously considered safe code patterns are being transformed into secret-dependent operations at the binary level.
Notable findings include issues in libraries that were formally verified to be free of such side channels~\cite{zinzindohoue2017hacl}. We discovered similar issues in all studied processor architectures, all compilers, and all optimization levels, including the recommended defaults. Our data indicates mainstream processor architectures, such as x86-64 and aarch64, to be less affected. We believe this is due to the developer's focus on such architectures.

Our findings demonstrate that currently applied defensive programming techniques are not always effective at mitigating side-channel attacks, are very fragile, and may be vulnerable to new releases of compilers. Where possible, we investigate the compiler implementation to pinpoint what is causing the secret dependencies to be re-introduced in otherwise constant-time-looking source code. 
Generally, we found that compilers understand certain arithmetic patterns and replace them with secret-dependent branches or memory accesses. 
Given these findings, our work highlights the need for a more comprehensive and better foundation for defensive programming techniques against side channels. 

\paragraph{Contributions}
\begin{itemize}
    \item We study how compilers break constant-time properties of source code and, therefore, introduce timing side-channels across major cryptographic libraries and architectures.
    \item We perform a large-scale study of 8 cryptographic libraries and 11 cryptographic algorithms on 6 different processor architectures, compiled by 2 off-the-shelf compilers with 9 and 13 respective compiler versions under 7 different optimization levels, leading to a total of \nrexperiments experiments.
    \item We demonstrate compiler-induced side channels in state-of-the-art hardened cryptographic libraries and find several such issues in multiple libraries. For \texttt{LLVM} and for specific non-constant-time transformations, we pinpoint the compiler pass that causes these issues.
    \item This paper highlights that defensive programming techniques for side-channel resistance are ineffective, incomplete, and fragile with modern compilers.
\end{itemize}

\paragraph{Disclosure}
We disclosed our findings to the respective developers of the libraries where we found any secret dependence (both compiler-induced and code-level issues): BearSSL, BoringSSL, Botan, HACL*, and Libsodium. Our disclosure included minimally reproducible examples of how certain compilers can break some parts of their implementation. The developers have acknowledged our findings, and two have provided fixes. %
The other library developers are currently working on a fix. The initial disclosure messages are provided in Appendix~\ref{app:disclosure}.

\section{Background}
\label{sec:background}

\paragraph{Side-channel Attacks}
Side-channel attacks have been studied extensively in the past years. 
There are various different types of such attacks measuring, e.g., power consumption~\cite{kocher1999differential}, changes in thermal emanations~\cite{masti2015thermal}, cache usage~\cite{liu2015last}, or the CPU branch predictor~\cite{aciiccmez2006predicting}. 
Physical side-channel attacks require the attacker to be in physical proximity of the device, e.g., to measure a power trace~\cite{kocher1999differential}. Digital side-channels can be executed remotely, e.g., by the attacker executing some other code on the same device~\cite{liu2015last}. Especially in the area of digital side channels, the attacker capabilities have been improved in the past few years with higher granularity~\cite{moghimi2019memjam,yarom2017cachebleed} and higher synchronization up to individual instructions~\cite{van2018nemesis}.
In this paper, we focus on such advanced digital side channels.

Variations exploited by side channels can be measured across the whole execution or at a finer granularity, e.g., functions, branches, and even single instructions~\cite{van2018nemesis}. 
These variations might be induced by the internal state of the CPU or some other component that is accessible to the attacker executing on the same platform. Usual examples are the shared cache~\cite{liu2015last} and the shared branch-predictor~\cite{aciiccmez2006predicting,lee2017inferring}. Unless the attacker can observe the leaky shared state directly, a common technique to exploit these variations is to convert them into timing variations. 

\paragraph{Constant-time Programming}
To counteract side-channel attacks, a crucial strategy in defensive programming is constant-time programming~\cite{aumasson2012cryptographycodingstandard,aumasson2019cryptocoding}. This methodology involves adhering to a set of rules that seek to eliminate any secret-dependent branching operation, memory access, and sometimes even instructions that are known to produce timing differences based on their arguments. A well-known example is the \texttt{select} function depicted in two different implementations in \cref{lst:example}, which returns either \texttt{x} or \texttt{y} based on a secret bit.
A non-constant-time-aware implementation of the \texttt{select} function (\Cref{lst:example:nonct}) would use an \texttt{if} statement to decide which value to return -- thus branching depending on a secret value.
Defensive programming proposes to replace the conditional statement with, e.g., the technique shown in \Cref{lst:example:ct}, that achieves the same result by using bit arithmetic without secret-dependent branches. Common techniques include using binary arithmetic to replace branches, removing early exit conditions of loops, and avoiding table look-ups indexed by secret data (e.g., s-boxes). We note that these transformations can be automated~\cite{borrello2021constantine,rane2015raccoon}.

\begin{figure}
\captionsetup{type=lstlisting}
\centering
\begin{subcaptionblock}{0.99\linewidth}
\begin{lstlisting}[language=myc]
uint32_t select(uint32_t x, uint32_t y, uint32_t bit) {
    if (bit) 
        return x;
    else 
        return y;
}
\end{lstlisting}
\caption{Not constant-time.}
\label{lst:example:nonct}
\end{subcaptionblock}

\begin{subcaptionblock}{0.99\linewidth}
\begin{lstlisting}[language=myc]
uint32_t ct_sel(uint32_t x, uint32_t y, uint32_t bit) {
    // Create a bitmask in constant-time: 
    // - 0x0 if bit == 0
    // - 0xFFFFFFFF if bit != 0
    uint32_t mask = create_bitmask(bit);
    // Select value to return in constant-time
    return (x & mask) | (y & ~mask);
}
\end{lstlisting}
\caption{Constant time.}
\label{lst:example:ct}
\end{subcaptionblock}
\caption{An example \texttt{select} function from~\cite{aumasson2019cryptocoding} that returns either \texttt{x} or \texttt{y} based on a secret bit. (a) is not constant-time due to branching on the secret value; (b) uses defensive programming techniques (bit arithmetic) to run in constant-time.}
\label{lst:example}
\end{figure}

\section{Problem Statement}
\label{sec:problemstatement}
Defensive programming techniques are often used to harden security-critical code to reduce the number of vulnerabilities and developer errors. 
Writing constant-time code is one such technique that seeks to mitigate side channels at the source code level~\cite{aumasson2012cryptographycodingstandard,aumasson2019cryptocoding}, and that experts rely on~\cite{bearssl}. 
These techniques aim to remove any secret-dependent memory accesses and control-flow operations, thus eliminating secret-dependent variability and limiting the side-channel susceptibility of these libraries. 
The intuition is that if no secret-dependent operations exist, then no side channel can exist either. 

Writing small assembly code by hand to achieve constant-time execution is common in cryptographic libraries. Usually, small critical routines are implemented in this fashion. However, this approach requires manual analysis for each architecture and thus needs a high level of effort and expertise for every supported architecture. 
Consequently, most projects use high-level source-code implementations with constant-time programming techniques. 
However, a compiler compiles high-level source code into a binary, which may optimize defensive programming techniques and re-introduce secret-dependent operations.

We aim to verify if such defensive programming techniques effectively mitigate side-channel attacks down to the binary level, considering today's compilers and available architectures, by analyzing arguably the most hardened libraries: cryptographic libraries.
Compilers and their different versions, optimization levels, and target architectures could affect the final binary. Therefore, we aim to verify whether the resulting binaries have any secret-dependent operations -- and whether the compiler introduced them.

Note that performing this analysis is challenging in practice because it requires being able to execute and profile binaries on many architectures in an automated fashion. Existing tools for analyzing such binaries are usually not portable and only work on one architecture, e.g., Intel Pin~\cite{luk2005pin,intelpin,weiser2018data} for Intel Processors, or LLVM-based approaches that only allow analyzing LLVM output~\cite{he2020ct}.

\paragraph{Scope}
\label{sec:attackermodel}
We focus our study on secret-dependent memory access and control flow operations that might be introduced by a compiler.\footnote{We do not consider instructions that exhibit variable timing based on their arguments.}
Since we focus on compiler-induced issues, we restrict the study to hardened side-channel free implementations. This makes it easy to isolate the impact of the compiler since if any issue is found in the binary, it must have been introduced by the compiler.
Cryptography libraries are good candidates due to their threat model: generally, they consider any secret dependency issue to be critical, and if those are absent, then the library itself can be considered side-channel free. We do not consider implementations that use blinding to hide timing variations. %

Not all secret-dependent operations are equal. Some might immediately leak the entire key, others only a single bit, and some might not be exploitable at all. We note that for such a study, the severity of the secret-dependent operations is irrelevant; they should not be present altogether if defensive programming techniques are followed. While the existence of secret-dependent operations does not imply that relevant information can be recovered, future attacks might improve this, e.g., by using a novel microarchitectural structure~\cite{ronen2018pseudo} or by increasing their granularity~\cite{albrecht2016lucky} -- this is why the prevalent defensive programming techniques for side-channel hardening aim at removing secret dependencies altogether. %

\section{Study Design}

\begin{figure*}[tbp]
	\centering
	\includegraphics[width=0.95\textwidth]{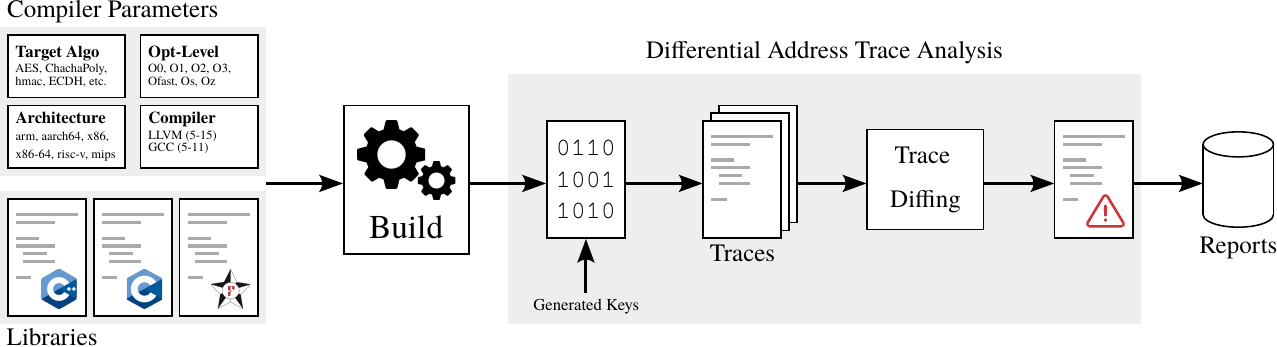}
	\caption{Pipeline used for our analysis. We use a dynamic analysis approach based on differential address trace analysis\cite{weiser2018data,he2020ct}.}
	\label{fig:pipeline}
\end{figure*}

We developed an analysis pipeline to study compiler-induced side-channel issues, as depicted in~\cref{fig:pipeline}. 
We first designed a build system that allows us to target different versions of different compilers and to specify lists of possible values for different parameters, e.g., target architecture and optimization level.
For each library we intended to study, we developed small applications that call the official APIs exposed by the libraries, e.g., \texttt{crypto\_box} in Libsodium. Both the library and the small applications are then compiled using the selected parameters.
Finally, the resulting binaries are analyzed using our tool, which finds secret-dependent operations and pinpoints their location in the analyzed library.

\subsection{Detecting Secret-Dependent Operations}
\label{sec:study:tool}

We developed a new tool~\cite{microsurf} based on emulating user-space binaries through Qiling~\cite{qiling} and Unicorn~\cite{unicorn}, which are based on QEMU.%
This emulation-based approach allows targeting any processor architecture and compiler, as it can process any final binary. At the same time, our tool can detect the same issues as similar binary instrumentation frameworks~\cite{weiser2018data}.
To fix potential non-determinism, our tool hooks a set of syscalls selected by hand to fix the randomness to a deterministic value that does not change across multiple executions.
Thus, by running in a fully deterministic environment, we posit that if we observe different execution traces for a given algorithm with two different secret keys, there is a secret dependency.
Our tool emulates the execution of the target binary a configurable number of times and records the execution trace of each run. We manually disable certain JIT optimizations in QEMU that impact the traces we want to collect.
Finally, our tool generates a report that states how many differences were found in traces, the functions that led to differences, and how often these were observed. Locating differences between multiple traces is trivial for memory-access traces but gets more involved for control flow. 
Just as in related work~\cite{weiser2018data,wichelmann2018microwalk,weiser2020big}, we compare the control-flow traces until the first difference is detected. Afterward, we look for a merge point where the traces re-align. This approach is quite compute-intensive as it requires a forward search with multiple traces and comparisons. We note that no further differences will be flagged between the trace difference and its merge point. Therefore, this approach only provides a lower bound for control-flow issues.

In general, our approach is similar to other trace-based dynamic analysis tools~\cite{weiser2018data,wichelmann2018microwalk} that compare sets of memory address traces for different secrets. The main difference is in the use of emulation, which allows us to fix non-determinism and enables us to analyze many different architectures and compilers. We note that we merely refine these trace-based approaches and inherit all their shortcomings and advantages. Nevertheless, a study like the one presented in this paper is impossible with existing tools.

\paragraph{Soundness} 
To verify the correctness of our tool, we reproduced all the results found in DATA~\cite{weiser2018data} for symmetric primitives in OpenSSL. In addition, we went through a significant number of flagged issues by hand to verify that they were real issues. We did not find any false positives.

\paragraph{Limitations}
We note that by fixing the randomness, we also fix the nonce and thus do not detect nonce leakage (that could also lead to compromised keys) as we remove any potential nonce dependencies.
Similarly, we do not detect any dependencies on the (fixed) plaintext or random numbers.
Our approach could analyze all these one at a time, but we defer this to future work.

Our study uses a dynamic analysis approach.
Thus, we can only find issues but not prove the absence thereof. We further arbitrarily set the number of runs to limit compute time. As a consequence, we might miss edge cases that only appear for some special conditions and would appear in a larger number of runs. Nevertheless, our findings demonstrate that we find relevant secret dependencies (c.f., \cref{sec:analysis:manual}), albeit we might have found even more with a higher number of runs. Our approach shows similarities with (and shares some limitations of) fuzzing systems that are also limited in their completeness and their execution time.

\subsection{Study Setup}

We conduct a large-scale analysis on two off-the-shelf compilers -- 9 versions of \texttt{GCC} and 14 versions of \texttt{LLVM} -- with 8 popular cryptographic libraries under 7 different optimization levels. This results in a total of 6608 distinct compiled binaries. We then analyze the binaries for up to 11 cryptographic primitives -- depending on the library -- resulting in \nrexperiments \textit{experiments}, i.e., compiled high-level functions. We then collect 8 traces for each experiment. %

We executed our study on eight machines equipped with an Intel Core i7-11700 and 32 GB of DRAM, and an AMD-based 192-core server with 1.5 TB of DRAM.
Our experiments took around seven days of computing time.

\paragraph{Targeted Cryptographic Libraries}
For this study, we selected 8 cryptographic libraries with varying claims of side-channel resistance: HACL*~\cite{zinzindohoue2017hacl}, Libsodium~\cite{libsodium}, Botan~\cite{botan}, BearSSL~\cite{bearssl}, BoringSSL~\cite{boringssl}, OpenSSL~\cite{openssl}, WolfSSL~\cite{wolfssl}, and MbedTLS~\cite{mbedtls}). We chose libraries that are widely used and/or highly regarded for their constant-time implementations. According to their documentation, all of these libraries strive to provide side-channel resistance at the source code level. HACL*~\cite{zinzindohoue2017hacl} claims to achieve this fully. We note that many of the side-channel-related claims are usually not made with respect to the resulting binary on all architectures using any compiler. Instead, side-channel resistance claims are usually only targeted at a limited set of architectures and compilers~\cite{libsodium} or for some intermediate representation~\cite{zinzindohoue2017hacl}. We analyzed the library's latest release or stable version as of March 2023:
\begin{itemize}
    \item HACL*/Project Everest~\cite{zinzindohoue2017hacl} (commit: f283af1)
    \item Libsodium~\cite{libsodium} (commit: 8cc84df)
    \item Botan~\cite{botan} (commit: 15dc32f)
    \item BearSSL~\cite{bearssl} (commit: 79c060e)
    \item BoringSSL~\cite{boringssl} (commit: 28f96c2)
    \item OpenSSL~\cite{openssl} (commit: a92271e)
    \item WolfSSL~\cite{wolfssl} (commit: 4fbd4fd)
    \item MbedTLS~\cite{mbedtls} (commit: 1873d3b)
\end{itemize}

\paragraph{Cryptographic Algorithms}
We also selected a set of algorithms to be investigated. The chosen algorithms should be supported by as many target libraries as possible and reflect a wide range of cryptographic primitives, from symmetric encryption and message authentication codes to cryptographic signatures and key exchanges.
We list the chosen algorithms and their respective support for constant-time implementations in the libraries in \cref{tab:algo}.

\begin{table*}[tbp]
    \centering
    \caption{The target cryptographic libraries and their respective support of constant-time cryptographic primitives. We use \yes and \no\xspace{} to indicate claimed constant-time and non-constant-time implementations respectively. Implementations that use blinding are indicated using \blind and handwritten assembly implementations using \asm.}
    \label{tab:algo}
    \newcommand{\tnport}{$\dagger$}
    \newcommand{\tabnote}{$^2$}
    \newcommand{\tn}{$^3$}
    \newcommand{\tc}{}
    \begin{tabular}{lcccccccc} \toprule
                         & HACL*          & Libsodium   & Botan & BearSSL & BoringSSL & OpenSSL & WolfSSL & MbedTLS \\ \midrule
        AES CBC          & \asm     & \no    & \yes   & \yes   & \yes      & \no     & \yes    & \no \\
        AES CTR          & \asm     & \no    & \yes   & \yes   & \yes      & \no     & \yes    & \no \\
        AES GCM          & \asm     & \asm   & \yes   & \yes   & \yes      & \no     & \yes    & \no \\
        Chacha20Poly1305 & \yes     & \yes   & \yes   & \yes   & \yes      & \yes    & \yes    & \yes \\
        HMAC SHA1        & \yes     & \no    & \yes   & \yes   & \yes      & \yes    & \yes    & \yes \\
        HMAC SHA256      & \yes     & \yes   & \yes   & \yes   & \yes      & \yes    & \yes    & \yes \\
        HMAC Blake2      & \yes     & \no    & \yes   & \no    & \yes      & \yes    & \no     & \no \\
        Curve25519       & \yes     & \yes   & \yes   & \yes   & \yes      & \yes    & \yes    & \blind \\
        ECDH P256        & \yes     & \no    & \blind & \yes   & \yes      & \blind  & \blind  & \blind \\
        ECDSA P256       & \yes     & \no    & \blind & \yes   & \yes      & \blind  & \blind  & \blind \\
        RSA              & \yes     & \no    & \blind & \yes   & \blind    & \blind  & \blind  & \blind \\
        \bottomrule
    \end{tabular}
\end{table*}

\paragraph{Compilers and Parameters}
We compiled the target libraries using one of 9 versions of GCC (from GCC 5 to GCC 13) or one of 14 versions of LLVM (from LLVM 5 to LLVM 18). %
We used MUSL as the underlying C standard library~\cite{musl}. Note that we do not trace MUSL itself and thus will not detect any secret dependency within it. We do this because we are interested in studying the cryptographic libraries themselves and not a specific \texttt{libc} environment in which they might run. Since LLVM does not ship with a compatible sysroot, we fall back on the sysroot provided by the GCC toolchain. 

We also investigated how optimization levels impact secret-dependent memory accesses and control flow. We target the following optimization levels: \texttt{-O0}, \texttt{-O1}, \texttt{-O2}, \texttt{-O3}, \texttt{-Ofast}, \texttt{-Os}, \texttt{-Oz}\footnote{The optimization level \texttt{-Oz} is only available in LLVM and is thus omitted for GCC.}. 

Finally, we also tested 6 processor architectures: x86-64, x86-i386, armv7, aarch64, RISC-V~\footnote{RISC-V is only supported starting from GCC version 10 and LLVM version 10 onwards}, and MIPS-32. We believe that this list represents a good mix of ISAs used for all kinds of computing systems. We also note that during the disclosure, the contacted library developers considered all these architectures as important.

All target libraries usually do not support such a large variety of compilers, optimization levels, and architectures. We developed some best-effort compilation scripts to be able to compile for all of these targets. However, some combinations of parameters either lead to a non-trivial error during compilation, a broken binary, or the resulting binary became so inefficient that our analysis ran out of memory on the smaller machines (mainly with \texttt{-O0}).

\paragraph{Removing Source-Code Issues}
Even though constant-time programming has been known for a long time, some of the targeted libraries still contain secret-dependent issues that are rooted in source code, e.g., AES implementations with lookup tables. Such issues are usually not removed by the compiler and thus also show up in our results. We target compiler-induced issues specifically and, therefore, need to filter any such issues. To achieve that, we maintain a manually curated list of functions or files that contain timing issues that originate from the source code and remove any issue that links back to this list. However, such filtering has its limitations: it does not always work due to compilers inlining functions or other optimizations that remove or modify the connection to the source code, and our manually curated list might miss some source-code issues and not remove them from our dataset. On the other hand, removing entire functions might also remove some compiler-induced issues within these functions. Nevertheless, for some libraries and compilers, it was impossible to filter results by function name as the compiler often removes this information. We manually verify that the other secret dependencies flagged by our tool stem from source code that seems to follow defensive programming guidelines for cryptographic code~\cite{aumasson2019cryptocoding}.

\paragraph{Automation}
To run such a wide-ranging study, we developed an experimentation framework~\cite{microsurfEval} based on Kubernetes. Our platform takes a set of parameters to evaluate as input, e.g., target the library BoringSSL with all supported compilers, all optimization levels, all architectures, and all algorithms. It then launches individual jobs for each combination (target algorithms are merged into the same job). Each job first sets up the required compiler and downloads the specified version of the target library. Then, it builds the library with the selected optimization level. Finally, our tool emulates the binary and collects traces for the selected cryptographic algorithms. Our tool then generates the reports and saves them to a shared network drive. %

\section{Analysis}

In this section, we quantitatively analyze how compiler parameters and libraries lead to compiler-induced secret-dependent operations. We first discuss how many secret-dependent operations were flagged and if there are differences between the processor architectures, compilers, compiler versions, and optimization levels. 
From the total of \nrexperiments experiments that we ran, 37,818 (84.8\%) were completed successfully. The remaining experiments failed either due to compiler errors or due to our system running out of memory during the address trace collection. Of the total number of 6,608 binaries, we compiled 5,716 successfully (86.5\%). Our experiments span 54 implementations of cryptographic primitives.

\subsection{Found Issues}
In the 5,716 binaries, we found at least one secret-dependent control-flow operation in 429 (7.5\%) and at least one secret-dependent memory access in 368 (6.4\%). %
In absolute numbers, we found 13,341 compiler-induced issues in the investigated libraries. We list the number of issues per library in \cref{tab:issuesperlib}. We note that the number of issues found is exaggerated because the same issue can appear in multiple places in a binary or trace (e.g., due to inlining a function). 
For example, in Botan's implementation of x25519, we found a total of 5,499 issues, 4,221 of which are found in binaries for MIPS. A brief manual analysis of the issues in Botan's x25519 implementation on MIPS shows the same issues popping up more than 100 times for every experiment. This wealth of issues can overshadow other more nuanced issues that do not appear that often. Therefore, in some of the following analyses, we simply consider the following four cases: (1) we did not find a secret dependency, (2) we found at least one control-flow secret dependency, (3) we found at least one secret-dependent memory access, and (4) we found both secret-dependent control-flow and memory accesses. 

\begin{table}[tbp]
    \centering
    \caption{Number of compiler-induced secret dependent operations found per target library across all parameters. Note that the number of reported issues may appear very large due to finding the same code patterns multiple times.}
    \label{tab:issuesperlib}
    \begin{tabular}{llr}
    \toprule
    Library & Crypto Primitive & Flagged Issues \\
    \midrule
    \multirow[t]{3}{*}{BearSSL} & rsa & 400 \\
     & ecdh-p256 & 61 \\
     & ecdsa & 46 \\
    \midrule
    \multirow[t]{2}{*}{BoringSSL} & ecdsa & 120 \\
     & ecdh-p256 & 104 \\
    \midrule
    \multirow[t]{8}{*}{Botan} & curve25519 & 5000 \\
     & chacha-poly1305 & 336 \\
     & aes-gcm & 33 \\
     & hmac-blake2 & 26 \\
     & hmac-sha1 & 26 \\
     & hmac-sha2 & 26 \\
     & aes-ctr & 26 \\
     & aes-cbc & 26 \\
    \midrule
    \multirow[t]{3}{*}{HACL*} & rsa & 6729 \\
     & ecdh-p256 & 244 \\
     & ecdsa & 136 \\
    \midrule
    Libsodium & curve25519 & 2 \\
    \midrule
    \multicolumn{2}{l}{\textbf{Total}} & \textbf{13,341} \\
    \bottomrule
    \end{tabular}
\end{table}

\subsection{Processor Architectures}
\begin{figure}[tbp]
    \centering
    \includegraphics[width=1.0\linewidth]{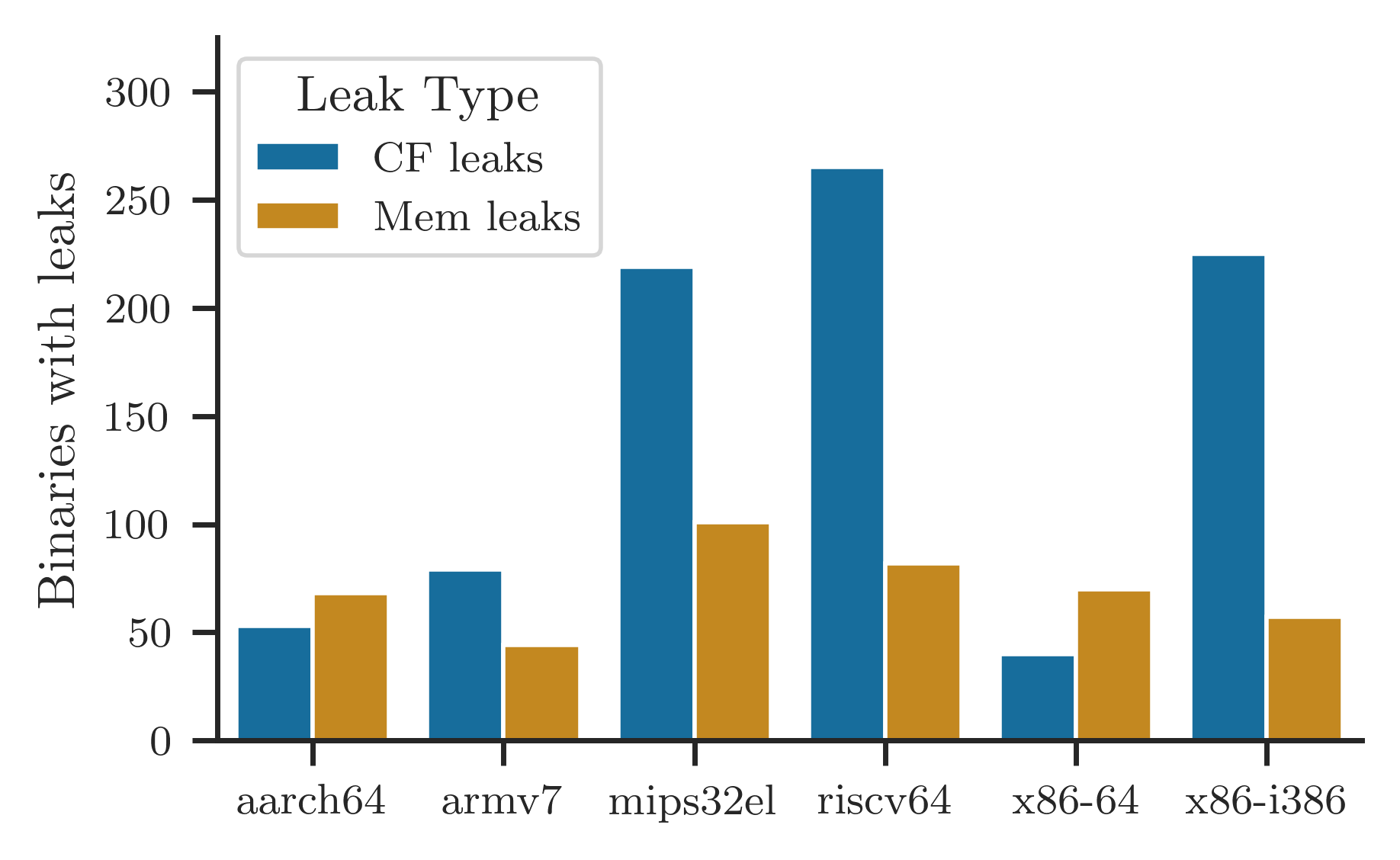}
    \caption{Binaries with at least one control-flow or memory-based secret-dependent operation by architecture. We observe more leaks in less popular processor architectures.}
    \label{fig:architectures}
\end{figure}

We depict all binaries with at least one control flow and memory accesses that were flagged to be secret dependent in \cref{fig:architectures}. In particular, there is a significant difference between processor architectures. We found more binaries with secret dependencies on less commonly used processor architectures, i.e., MIPS, RISC-V, and x86-i386. This is especially notable for RISC-V, which is only supported in 8 out of 17 compiler versions and thus has less than half the experiments of other architectures. Nevertheless, we found more binaries with at least one secret-dependent operation in RISC-V compared to x86-64, armv7, and aarch64. More mainstream processor architectures seem to be less affected but not entirely free of compiler-induced issues. 

\subsection{Compilers and Optimization Levels}

\begin{figure*}[tbp]
    \centering
    \begin{subcaptionblock}{0.485\linewidth}
        \centering
        \includegraphics[width=1\linewidth]{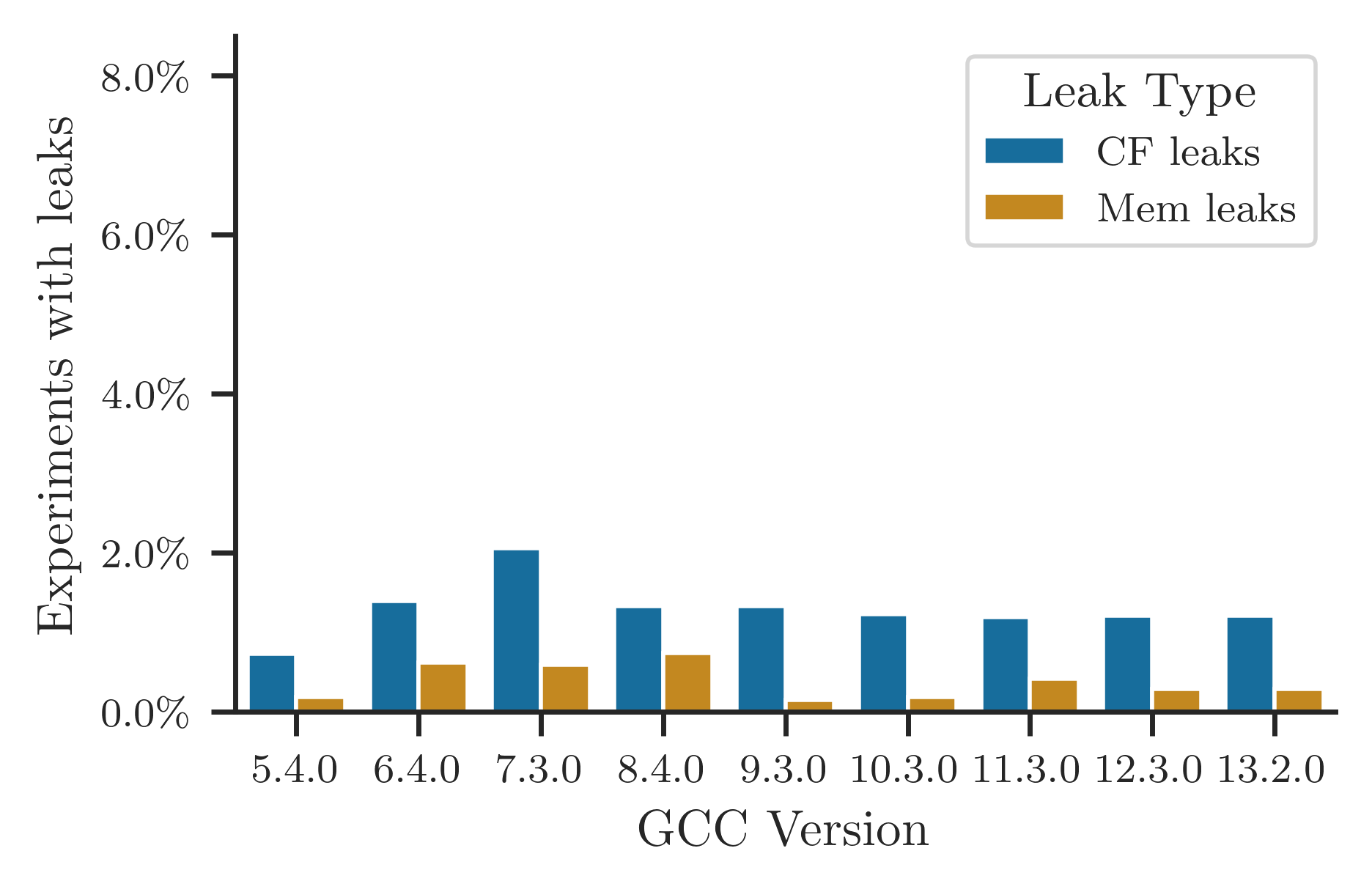}
    \end{subcaptionblock}
    \begin{subcaptionblock}{0.485\linewidth}
        \centering
        \includegraphics[width=1\linewidth]{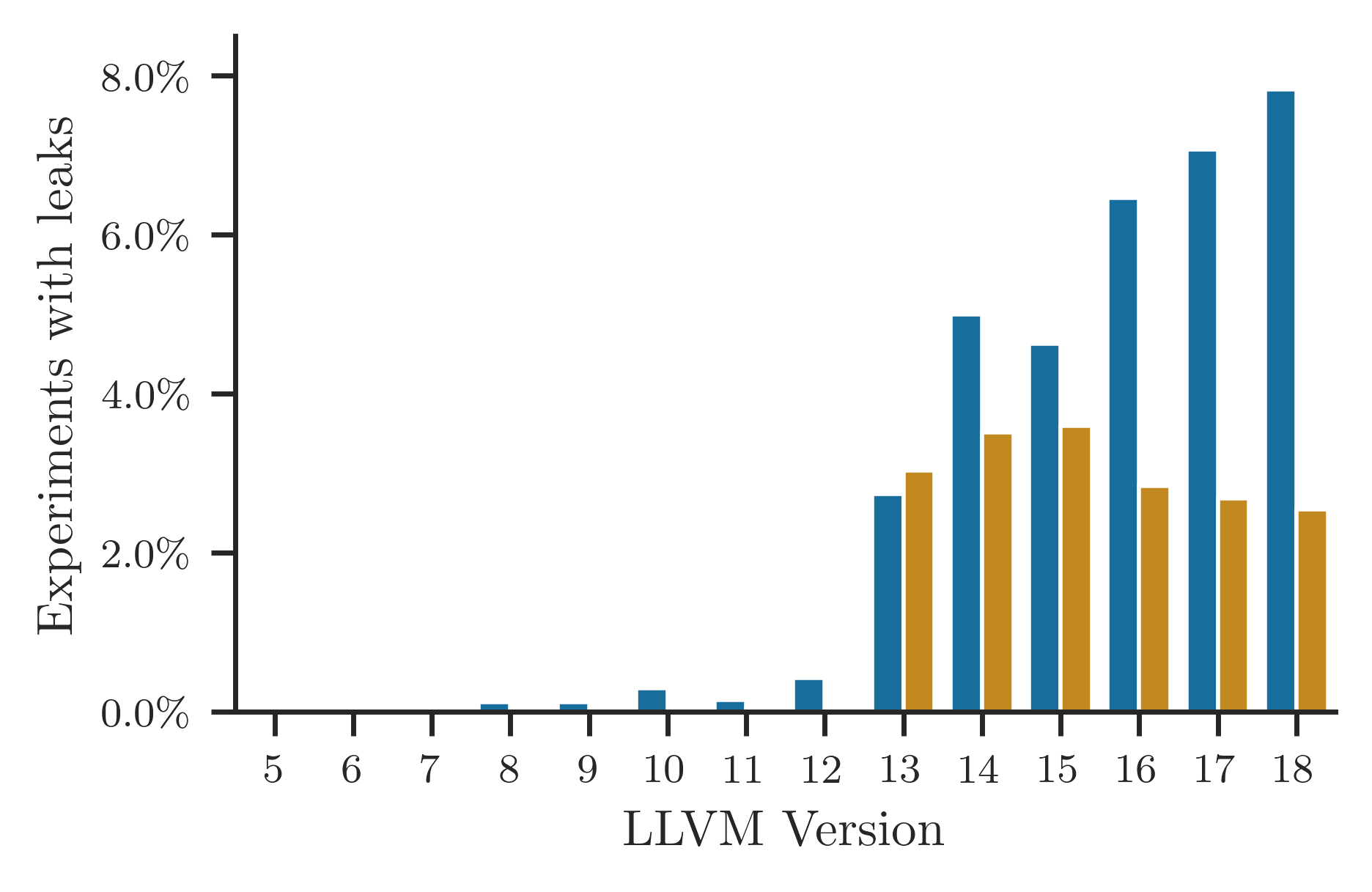}
    \end{subcaptionblock}
    
    \begin{subcaptionblock}{0.485\linewidth}
        \centering
        \includegraphics[width=1\linewidth]{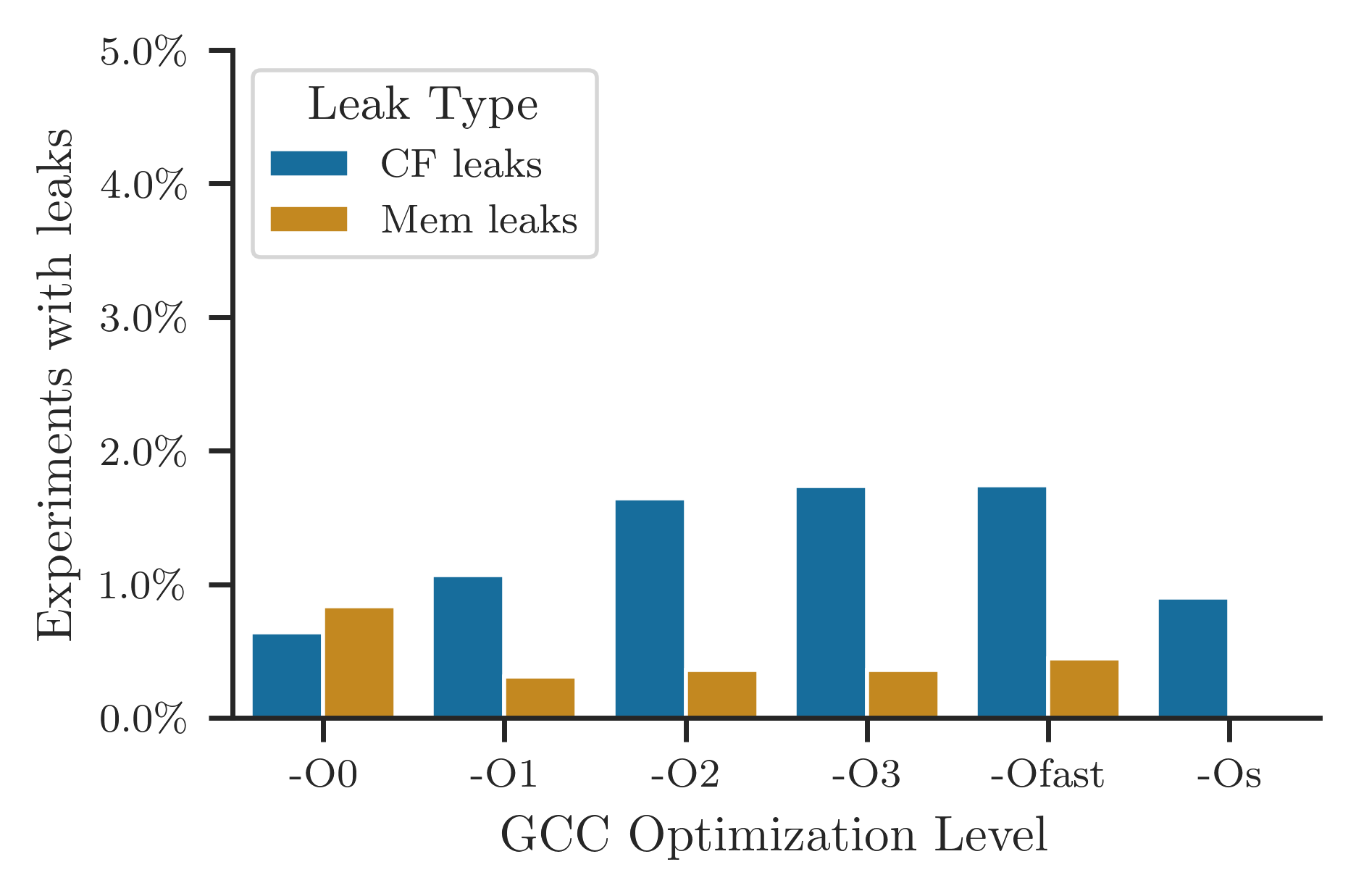}
    \end{subcaptionblock}
    \begin{subcaptionblock}{0.485\linewidth}
        \centering
        \includegraphics[width=1\linewidth]{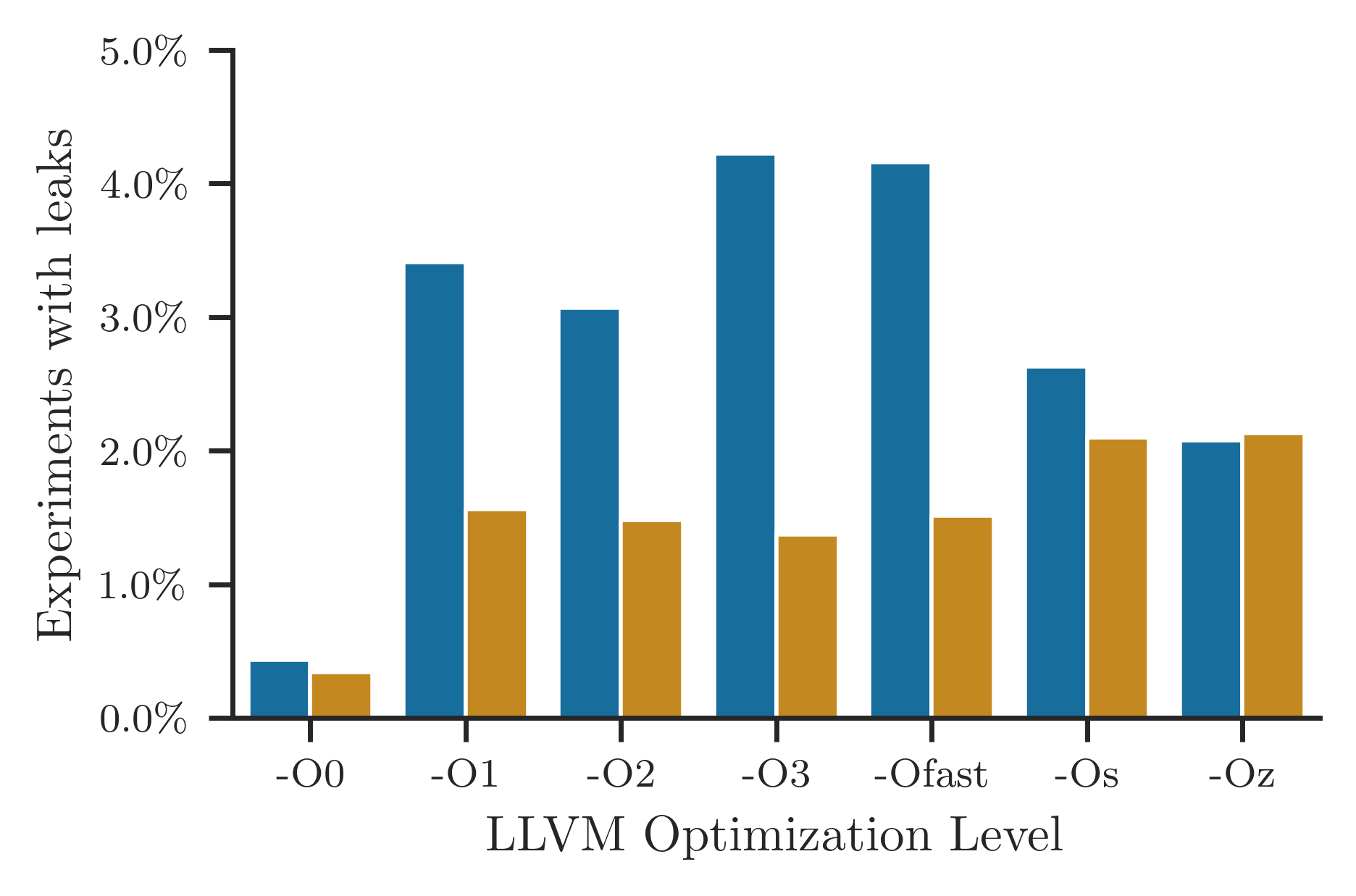}
    \end{subcaptionblock}
    \caption{Experiments with at least one secret-dependent operation per compiler, respective versions, and optimization level.}
    \label{fig:compilers}
\end{figure*}

We depict the number of binaries in which some issues were flagged according to the compiler and the version that produced it in \cref{fig:compilers}. Interestingly, older compilers seem to introduce fewer issues. We believe this is due to more aggressive optimizations in newer compilers, potentially undoing defensive programming techniques. Some of these optimizations are discussed in the following section (\cref{sec:analysis:manual}), where we manually analyze individual binaries.

We found binaries with issues in all optimization levels as depicted in \cref{fig:compilers}. We observe notable drops for LLVM's and GCC's \texttt{-O0}. We believe that fewer enabled optimizations also lead to fewer compiler-induced issues. We discuss in \cref{sec:analysis:manual:discussion} that optimizations are one of the reasons compilers introduce side channels; thus fewer optimizations could mean defensive programming techniques are less likely to get interfered with. However, we note that this cannot be considered a solution to compiler-induced issues as we still observe some issues with \texttt{-O0}. We also note that the \texttt{-O0} results may be misleading since we could not analyze some algorithms in some libraries with \texttt{-O0} as our machines ran out of memory. Therefore, the experiments with \texttt{-O0} are not as thorough as the other optimization levels. 
We also observe a reduction in flagged issues with \texttt{-Os}. It is unclear why these differences exist.

\section{Notable Findings}
\label{sec:analysis:manual}

We perform an in-depth manual analysis for a subset of the found secret-dependent operations. 
Here, we discuss some notable findings by showing the portable source code snippet and the corresponding binary introducing the secret dependency. If possible, we also briefly analyze the underlying mechanism in the compiler that introduces the secret dependency. We only discuss the simplest examples or those that we can minimally reproduce in this section. %

\subsection{HACL*}
\paragraph{Secp256}
In HACL*, we discover that the compiler can introduce secret-dependent control flow into the implementation of the elliptic curve secp256. This secret dependency is executed in HACL*'s implementation of ECDH-p256 and ECDSA, as both rely on the secp256 implementation.
The code snippet that leads to the secret dependency can be found in \Cref{lst:haclstar}. 
The intended functionality of the function \texttt{cmovznz} is a conditional move: it moves either $x$ or $y$ to $r$ depending on a secret $condition$ (in constant time). 
The function achieves this by first computing a $mask$ of either all 1's or all 0's depending on the condition (line 4 in \cref{lst:haclstar:code}). 
Afterwards, $r$ is set to $(y \wedge mask) \vee (x \wedge \neg mask)$. Thus setting $r$ to either $x$ or $y$ depending on the $mask$ and implicitly the $condition$ (line 5 and 6 in \cref{lst:haclstar:code}). One would expect this to be compiled into some binary AND and OR operations. 
However, some versions of LLVM understand that this is the same as a branch and replace the bit-wise operations. On MIPS and armv7, LLVM replaces the bitmask with a direct calculation of the address of the selected value to be stored in \texttt{r}. We depict an example resulting armv7 binary in \cref{lst:haclstar:asm}. On both MIPS and armv7 we observed this behavior for versions 14 and 15 and optimization levels \texttt{-Os} and \texttt{-Oz}. On RISC-V and x86-i386, LLVM replaces the bitmask operations with four control flow instructions. We observe this behavior on LLVM versions 13, 14, and 15 on all optimization levels besides \texttt{-O0}. 

Further investigation shows that LLVM understands the bitmask generation and arithmetic logic and replaces it with a \texttt{select} IR instruction with the semantic of assigning a variable to one of the parameters depending on the input condition. The culprit of this optimization is a pass called \texttt{InstCombine}, which performs optimizations by combining and simplifying arithmetic operations. In this case, \texttt{InstCombine} replaces the mask generation and the subsequent bitwise AND and ORs with a \texttt{select}. According to the LLVM documentation~\cite{llvmdoc}, this is a potential optimization to allow the backend to decide how to instantiate the \texttt{select} IR instruction based on the processor architecture. For example, on x86-64, \texttt{select} is sometimes instantiated with a \texttt{cmov} instruction. 

\begin{figure*}
\captionsetup{type=lstlisting}
\centering
\begin{subcaptionblock}{0.477\textwidth}
\begin{lstlisting}[language=myc]
void cmovznz4(uint32_t condition, uint32_t *x, 
              uint32_t *y, uint32_t *r) {
    uint32_t mask = ~calc_mask(condition);
    r[0] = (y[0] & mask) | (x[0] & ~mask);
}
\end{lstlisting}
\caption{Source code snippet.}
\label{lst:haclstar:code}
\end{subcaptionblock}
\hfill
\begin{subcaptionblock}{0.477\textwidth}
\begin{lstlisting}[language=asm]
cmovznz4:
    cmp     r0, #0
    movne   r1, r2
    ldr     r0, [r1]
    str     r0, [r3]
    bx      lr
\end{lstlisting}
\caption{armv7 assembly (LLVM 15, \texttt{-Os}).}
\label{lst:haclstar:asm}
\end{subcaptionblock}
\caption{Simplified snippet used in HACL*'s implementation of elliptic curve P256. This function implements a conditional move, i.e., it moves either \texttt{x} or \texttt{y} to \texttt{r} depending on \texttt{condition}. The function \texttt{calc\_mask} creates a bitmask based on $condition$ with either all 1s or all 0s. LLVM replaces the bitmask arithmetic with a \texttt{movne} to compute the address followed by secret-dependent memory accesses in armv7 and MIPS.  In RISC-V and x86-i386, the bitmask arithmetic gets replaced with a secret-dependent branch instead.}
\label{lst:haclstar}
\end{figure*}

\subsection{Botan}

\paragraph{AES-GCM}
We discover compiler-induced secret-dependent control flow in Botan's implementation of GHASH used in AES-GCM.
\Cref{lst:botan:aesgcm} shows a simplified snippet of how Botan precomputes the H-values for the key schedule of GHASH in a supposedly purely arithmetic way. However, instead of just performing the XOR operation with \texttt{carry}, LLVM 15 replaces this with a branch on RISC-V: If \texttt{carry} is 0, then jump over the XOR. %

\begin{figure*}
\captionsetup{type=lstlisting}
\centering
\begin{subcaptionblock}{0.477\textwidth}
\begin{lstlisting}[language=myc]
uint64_t precompute(uint64_t H0, uint64_t H1) {
    const uint64_t R = 0xE100000000000000;
    const uint64_t carry = R * (H1 & 1);
    return (H0 >> 1) ^ carry;
}
\end{lstlisting}
\caption{Simplified source code.}
\label{lst:botan:aesgcm:code}
\end{subcaptionblock}
\hfill
\begin{subcaptionblock}{0.477\textwidth}
\begin{lstlisting}[language=asm]
precompute:
    andi    %
    srli    %
    beqz    %
    li      %
    slli    %
    xor     %
.LBB0_2:
    ret
\end{lstlisting}
\caption{RISC-V assembly (LLVM 15, \texttt{-O2}).}
\label{lst:botan:aesgcm:asm}
\end{subcaptionblock}
\caption{Simplified source code snippet found in Botan's implementation of GHASH used in AES-GCM. This function is a part of the key schedule. The compiler produces a binary that skips the addition if the carry is 0.}
\label{lst:botan:aesgcm}
\end{figure*}

\paragraph{128-bit Custom Type}
We discover a similar compiler-induced secret-dependent operation in the implementation of a custom 128-bit wide type in Botan used in Chacha-Poly1305 and x25519. For platforms that do not natively support 128-bit integers, Botan uses a 128-bit software implementation that uses two 64-bit words. Botan implements the addition operator by first adding the lower half, calculating the carry, and then adding the upper half and the carry. We depict Botan's implementation in \cref{lst:botan:chachapoly}. Similar to the previous issue in Botan's GHASH implementation, binaries by some compilers first check the carry and skip the second addition if the carry is not set. We observe this behavior for GCC on MIPS and x86-i386 under all optimization levels (including no optimization \texttt{-O0}). Interestingly, the compiler can generate both secret-dependent control flow and memory accesses at the same time. For example, using GCC 11.3.0 on MIPS with \texttt{-O2} produces a binary with memory and control-flow secret dependencies. We observe this operator getting called in Botan's x25519 and chacha-poly1305 implementation.

\begin{figure}
\captionsetup{type=lstlisting}
\centering
\begin{lstlisting}[language=myc]
uint128_t operator+=(uint128_t &x) {
    l += x.l; h += x.h;
    uint64_t carry = (l < x.l);
    h += carry;
    return *this;
}
\end{lstlisting}
\caption{Simplified source code snippet found in Botan's implementation of a 128-bit type used in chacha-poly1305. The compiled binary skips the carry addition if the carry is 0.}
\label{lst:botan:chachapoly}
\end{figure}

\subsection{BearSSL}

\paragraph{Secp256}
BearSSL's implementation of the elliptic curve secp256 contains a small routine to check a scalar; a simplified version is shown in \cref{lst:bearssl:p256}. In this function, a scalar is checked to be below the curve order. This is achieved by comparing the scalar byte-by-byte to the curve order (\texttt{P256\_N}). Recent versions of LLVM for RISC-V replace the bitmask arithmetic operations with a secret-dependent branch. 

\begin{figure}
\captionsetup{type=lstlisting}
\begin{lstlisting}[language=myc]
uint32_t check_scalar(const unsigned char *k) {
    uint32_t c = 0;
    for (int u = 0; u < 32; u ++) {
        c |= CMP(k[u], P256_N[u]);
    }
    return c;
}
\end{lstlisting}
\caption{Simplified \texttt{check-scalar} routine in BearSSL's implementation of secp256. This routine checks if the key is lower than the curve order. The function \texttt{CMP} compares two numbers and returns -1, 0, 1 if the first argument is smaller, equal, or larger using binary arithmetic operations. Recent LLVM versions on RISC-V detect the bitmask arithmetic and replace it with a secret-dependent branch.}
\label{lst:bearssl:p256}
\end{figure}

\paragraph{Modular Exponentiation}
We discover another compiler-induced secret-dependent operation in BearSSL's implementation of modular exponentiation. The source code snippet and the corresponding x86-64 assembly can be found in \cref{lst:bearssl:modexp}. The desired functionality of the snippet is similar to the \texttt{cmovnz} snippet described above for HACL*. The snippet either sets \texttt{x} to \texttt{y} or to \texttt{z} depending on the condition. \texttt{EQ} computes a mask of either all 0's or all 1's depending if \texttt{condition == 0}. LLVM on Aarch64 and x86-64 replace the mask computation with a conditional \texttt{mov} on x86-64 and a conditional \texttt{select} on aarch64, respectively. However, the compiler does not use this instruction to set the \texttt{x} variable directly. It uses it to compute an address, which is then loaded and stored in \texttt{x}. This leads to a secret-dependent memory access operation which leaks the condition as either the memory location of \texttt{y} or \texttt{z} is read and written into \texttt{x}. LLVM performs a part of this optimization in the \texttt{InstCombinePass} where the bitmask computation is replaced with a \texttt{select} IR instruction (similar to HACL*). In the backend, the compiler then replaces the select with a \texttt{cmov} or \texttt{csel} instruction, which computes the address for the secret-dependent memory access.

\begin{figure*}
\captionsetup{type=lstlisting}
\centering
\begin{subcaptionblock}{0.477\textwidth}
\begin{lstlisting}[language=myc]
uint64_t mask1 = -(uint64_t)EQ(condition, 0);
uint64_t mask2 = ~mask1;
for (int i = 0; i < len; i ++) {
    x[i] = (mask1 & y[i]) | (mask2 & z[i]);
}
\end{lstlisting}
\caption{Source code snippet.}
\end{subcaptionblock}
\hfill
\begin{subcaptionblock}{0.477\textwidth}
\begin{lstlisting}[language=asm]
test    %
lea     %
lea     %
cmove   %
mov     %
mov     qword ptr [%
\end{lstlisting}
\caption{x86-64 assembly (LLVM 15, \texttt{-Os}).}
\end{subcaptionblock}
\caption{Simplified source code snippet in BearSSL. Similar code patterns are used in its modular exponentiation implementations used in RSA, ECDH-P256, and ECDSA. Recent versions of LLVM on x86-64 replace the mask with a \texttt{cmov} instruction to create the address that is used to write a value into \texttt{x[i]}. This thus leaks the condition as the address of \texttt{y[i]} is different to \texttt{z[i]}. On Aarch64, LLVM inserts \texttt{csel} instead of \texttt{cmov}.}
\label{lst:bearssl:modexp}
\end{figure*}

\subsection{BoringSSL}
\paragraph{Complex Branching Condition}
Recent LLVM compilers may split a branch with complex conditions into multiple branches with parts of the condition in a snippet found in BoringSSL. This behavior creates a small leak because which branch is taken leaks which part of the condition was true or false. We observed this behavior in the point-add routing of P256 on armv7, MIPS, and x86-i386. A snippet that shows the culprit branch is depicted in \cref{lst:boringssl:branch}. %

\begin{figure}
\captionsetup{type=lstlisting}
    \centering
    \begin{lstlisting}[language=myc]
uint32_t is_nontriv_double = ct_is_zero(xneq | yneq) & 
                            ~ct_is_zero(z1nz) &
                            ~ct_is_zero(z2nz);
if (is_nontriv_double) {
    fiat_p256_point_double(x3, y3, z3, x1, y1, z1);
}\end{lstlisting}
    \caption{Simplified source code snippet in BoringSSL. Recent versions of LLVM on armv7, MIPS, and x86-i386 insert two distinct branching instructions instead of a single branch. Which of the two branching instructions is taken leaks information to an attacker and depends on the secret.}
\label{lst:boringssl:branch}
\end{figure}

\section{Problematic Compiler Internals}
\label{sec:analysis:manual:discussion}
We categorize the found compiler-induced secret-dependent operations into three types: replacing bitmask arithmetic with branches, arithmetic shortcuts to skip computation, and splitting complex branching conditions into multiple branches. The first issue was previously studied in toy examples~\cite{simon2018you}, but to the best of our knowledge, the latter two have not yet been discussed. We analyze how and why these optimizations are introduced in modern compilers. We note that this list is not comprehensive, as more compiler-induced secret dependencies might appear in future compilers. Since these code patterns are also used in other libraries without any introduced secret-dependent operations, we analyze whether these patterns can be used safely.

\paragraph{Simplification of Bitmask Arithmetic}
LLVM seems to be very advanced in understanding and optimizing bitwise arithmetic operations, as we found most such issues in binaries produced by LLVM. However, we also found some simplifications in binaries compiled by GCC, albeit less than with LLVM. In the following, we focus on LLVM as it optimizes bitmask arithmetic more aggressively. 

Compilers do numerous optimization passes on the input program. In the context of bitmask arithmetic, LLVM contains an optimization pass that tries to reduce redundant instructions called \texttt{instcombine}~\cite{llvmdoc}. This pass tries to perform algebraic simplifications by employing a big recursive pattern-matching engine. In the observed cases, the \texttt{instcombine} pass replaces the mask generation and the subsequent bitwise AND and ORs with a \texttt{select} LLVM instruction. The LLVM compiler instantiates the \texttt{select} instruction in the ISA-specific backend based on the processor architecture. For example, on x86-64, \texttt{select} is sometimes instantiated with a \texttt{cmov} instruction, and on RISC-V and MIPS with conditional branches. Sometimes, the ISA-specific backend removes the \texttt{select} instruction altogether and does not introduce secret-dependent operations. 

Since bitmask arithmetic is a popular building block in constant-time cryptographic implementations, we also investigate how other libraries leverage bitmasks without producing secret-dependent operations. Many libraries use the same implementation of various cryptographic algorithms, such as poly1305-donna~\cite{poly1305donna}. This implementation of poly1305 contains a bitmask arithmetic pattern in the function \texttt{poly1305\_finish}. In our experiments, no compiler produced a binary with secret-dependent operations for this implementation. We do not know if the current source code pattern is deliberately chosen to avoid secret-dependent operations or if this is just a lucky coincidence. For now, this implementation seems not to produce secret-dependent operations with the evaluated compilers. To test the robustness of code patterns using bitmask arithmetic, we randomly tried innocent source code changes and observed that compilers once again introduce secret-dependencies\footnote{E.g., we only moved the select bitmask part to its own function.}.
This highlights that bitmask arithmetic is not very robust and may unexpectedly produce secret-dependent operations. We did not find a way to ascertain if a specific arithmetic source code pattern is safe just by looking at the source code.

\paragraph{Arithmetic Shortcuts}
Arithmetic shortcuts use branches to skip further computation (c.f., \Cref{lst:botan:aesgcm}). We observed this behavior with both GCC and LLVM. Further manual experiments with source code patterns show that this optimization is mainly performed in the ISA-specific backend of the two compilers. We believe the issues found in this study to be misguided optimizations. Skipping an XOR instruction will be slower than just performing the XOR as it is a very fast instruction. In the instances found in this study, the compiler's frontend injects the secret-dependent branch via a \texttt{select} IR instruction, similar to the bitmask arithmetic optimization above. Since all architectures share the same frontend in LLVM, they all have a \texttt{select} instruction in their IR before the backend optimizations. Some backends for less popular architectures, such as MIPS and RISC-V, produce inefficient binaries that use branches to skip an XOR operation in some LLVM versions. Luckily, backends for the other four architectures remove the secret-dependent \texttt{select} again and just perform the XOR. 

To further experiment with this behavior, we manually create a code snippet that would greatly benefit from optimization: we created a function with a boolean input that allows a division operation to be skipped if it is true. Advanced compilers know that division is an expensive operation and, therefore, produce a binary that skips the division if possible (e.g., when the input is \texttt{true}). We managed to reproduce this with LLVM in arguably the most popular architecture, x86-64, where an ISA-specific pass introduces a branch that skips the division. 

In summary, branches to skip expensive computations are a common optimization in both compilers. They are instrumental in optimizing performance in cases where the expensive operation is often unnecessary and can be skipped. We observed these optimizations in multiple places in modern compilers: in the shared frontend and the architecture-specific backend of the compiler. Current issues that we found seem to be misguided and can be fixed with newer compilers with better optimizations. However, we found other patterns that did not occur in the surveyed libraries where secret-dependent branches would be injected. Therefore, developers should be careful in cases where an expensive operation could be skipped for 50\% of the inputs.

\paragraph{Complex Branch Conditions}
In some cases, compilers can split branches with complex conditions into multiple branching instructions, each with a part of the condition (c.f., \cref{lst:boringssl:branch}). This behavior allows an adversary to leak which branching instruction was taken and thus learn some information about the complex condition. The developer might assume that the condition is checked atomically, but the compiler gives no such guarantee. For the same architecture and compilation flags, sometimes a source code pattern is safe. However, just slight tweaks to the code (still adhering to the same pattern) lead to the compiler introducing secret-dependent operations. Therefore, avoiding such compiler-induced secret dependencies is very hard without binary analysis. However, we have only rarely found this behavior and only in LLVM.

\section{Takeaways}

\paragraph{Defensive programming techniques can be ineffective}
As we have shown in the previous sections, cryptographic libraries often contain compiler-induced secret dependencies even though they apply current defensive programming techniques. The manual analysis of the flagged issues indicates that the issues that we find are not due to developer error but rather unexpected transformations by the compiler. We thus conclude that the current defensive programming techniques are insufficient. It remains unclear if the rules are incomplete, i.e., a new specific rule would lead to safe source code again, or if they are fundamentally broken with modern compilers. We defer the exploration of such new rules to future work. 

\paragraph{Constant-time source code is brittle}
As seen in our analysis of findings in \cref{sec:analysis:manual}, compilers apply very smart simplifications and optimizations to source code.
However, we also demonstrated that some of these seem to introduce side-channel vulnerabilities in some implementations. The fact that only some implementations suffer from this suggests that compilers currently only optimize certain code patterns in ``dangerous'' ways.
Avoiding such patterns is not a definite solution, though: the same code patterns that, for now, seem to be fine based on our analysis might be broken by the next generation of compilers. 
For example, \texttt{LLVM} started to aggressively simplify bit-wise mask arithmetic to \texttt{select} IR instructions starting from version 12.
Our results indeed show a spike in the number of found vulnerabilities from that version in \cref{fig:compilers}. For now, \texttt{LLVM} only simplifies some types of bitmasks dangerously, and it does not recognize others -- future improvements might change this and lead to even more such issues.

Recommendations for libraries are that new releases should come with a list of tested target architectures, operating systems, compilers, and flags. 
Then, users of a library can at least use a compiler that is believed to maintain the constant-time properties of the library. Nevertheless, our work demonstrates how brittle the current state of constant-time high-level source code has become.

\paragraph{The Case for Continuous Testing}
Some of the issues that we discovered are produced by compilers that are up to 10 years old. This is a significant time period, and it is unclear why such behavior was not flagged before. We believe that rigorous continuous testing on as many processor architectures and compilers as possible should be the standard for cryptographic libraries. A recent survey~\cite{jancar2022they} has asked developers of cryptographic libraries about the tools they use and highlights that most libraries still lack such infrastructure. Our findings also confirm this. We approached the maintainers of the libraries targeted in this paper to offer our help if they want to add our tool to their infrastructure. Some of them were receptive to this idea, and we are currently evaluating if this is feasible.

\section{Defenses}

\paragraph{Principled Defenses}
There are at least two principled defenses that eliminate all compiler-induced secret dependencies: formal analysis of the \emph{binary} or using a compiler that guarantees not to include any issues. 

Formal verification of assembly is used on the predominant processor architectures~\cite{bond2017vale} only because it requires a high degree of effort and know-how of the targeted processor architecture.
Nevertheless, such code is guaranteed to remain constant-time as it does not go through any other transformation.

The second principled defense is using a compiler that guarantees no such induced issues. Ideally, the compiler comes with formal guarantees that the behavior of the source code in constant time will not be changed, such as CompCert~\cite{barthe2014system,barthe2019formal}. However, such compilers are rarely used in practice as they lag behind commodity compilers in supported architectures and performance.
There have also been many proposals to add passes to commodity compilers or to annotate functions or variables to disable certain optimizations~\cite{cauligi2019fact}. However, these efforts have not yet had the desired effect on mainstream compilers and have not been included upstream.

Compilers that transform secret-dependent code into constant-time binaries~\cite{borrello2021constantine,rane2015raccoon} usually transform general-purpose code into a linear binary that executes all branches no matter the input. These approaches have been shown to produce side-channel free binaries~\cite{borrello2021constantine}, but they are usually research artifacts and, as such, not yet usable for production. %

\paragraph{Certification}
Another option is certification: Cryptographic libraries could use tools to certify any binary produced by a commodity compiler, thus guaranteeing that the binary does not contain secret-dependent operations. Formal or symbolic tools could prove the absence of secret-dependent operations in binaries.
We note that the approach used in this paper cannot be used to provide this guarantee, as we only find such issues but cannot prove the absence thereof. Certifying binaries notably shifts the release procedure of major libraries as they should no longer release the source code to be compiled but rather multiple binaries for different architectures and operating systems. This would require a significant effort for every single release. Obviously, an important aspect of these libraries is the fact that they are open source, thus binary certification would be most beneficial in conjunction with approaches akin to reproducible builds.

\paragraph{Increased Testing}
Some libraries already perform in-depth testing prior to a new release. For instance, libsodium~\cite{libsodium} performs tests and analysis for more than 20 target platforms: from WebAssembly on V8, various compilers for x86-64 Ubuntu and macOS, to Debian on SPARC. As we did not find many issues in libsodium, this rigorous testing seems to have a beneficial effect. However, we want to note that libsodium is the smallest library that we test with the most limited support for cryptographic primitives. For example, libsodium does not support the elliptic curve secp256; instead, they do all of their elliptic curve operations on curve25519. Therefore, it may be unfair to compare to larger libraries with support for more cryptographic primitives. 

We remark that increased testing does not provide any formal guarantee of the binary compared to certification. Nevertheless, rigorous testing can help mitigate compiler-induced issues and detect them before they can do any harm.

\section{Related Work}

This paper presents contributions in two main directions: i) a new tool that allows testing for side-channels in many architectures from a single platform, and ii) a study of dangerous compiler transformations in constant-time libraries. We stress that the literature is rich in terms of related tools, and while its novelty is nuanced, as we describe below, its main advantage is that it enables performing such a study.

\paragraph{Side-Channel Detection Tools}
Numerous related tools detect side channels in various targets, from static analysis of higher-level source code to statistical tests of the final binaries. For an extensive survey of such tools, we refer to ~\cite{barbosa2021sok} and ~\cite{jancar2022they}.

Statistical test tools collect numerous black-box measurements and use statistical tools to interpret the measurements. For example, dudect~\cite{reparaz2017dudect} uses Welch's t-test to check for information leakage. While any leaks found by such tools are usually immediately exploitable, they also miss a significant number of leaks as they only consider their target as a black box. In practice, side-channel attacks have become much more powerful with improved granularity and synchronization~\cite{van2018nemesis,moghimi2019memjam}.

Symbolic trace-based tools~\cite{doychev2015cacheaudit} use symbolic execution to evaluate all paths of a target. CacheAudit~\cite{doychev2015cacheaudit}, one of the most well-known tools in the category, uses these traces to provide an upper bound for the leakage. It is easy to reason about a zero upper bound, but any other value is hard to interpret, as it does not help to pinpoint the issue. However, these tools are usually tailored to one processor architecture and cannot easily be extended to others.

Formal analysis tools provide strong guarantees as they can prove the absence of issues in a target~\cite{zinzindohoue2017hacl,almeida2013formal,barthe2014system}. However, these tools usually only reason on high-level source code or some intermediate representation. It is unclear if and how these guarantees translate to binaries\footnote{If a formally proven compiler~\cite{barthe2018secure} is used, then the guarantees should translate down to the binary.}.

Dynamic tools enable the analysis of the actual execution of the final binary. For example, ctgrind~\cite{langley2010ctgrind} uses Valgrind to trace hand-annotated secret memory regions and report any secret-dependent access to it. The closest related tools to the one developed for our study are trace-based dynamic analysis tools~\cite{he2020ct,weiser2020big,wichelmann2018microwalk}. Trace-based approaches record a number of traces for different secret inputs and then compare the traces to identify secret-dependent operations. These approaches have various advantages: they are relatively performant, reason about the final binary, and can precisely pinpoint the issues. 
However, they are incomplete: they cannot guarantee the absence of leaks. A final limitation of these tools is their limited portability, as they usually rely on binary instrumentation frameworks that are not portable. As such, these tools usually only support a single processor architecture~\cite{weiser2018data}.

\paragraph{Automatically Fixing Secret Dependencies}
This paper focuses on libraries that are already hardened against all side-channel attacks. However, the vast majority of programs are not hardened. Various approaches that transform non-hardened code to a binary that contains no secret dependencies exist~\cite{rane2015raccoon,borrello2021constantine}. However, we note that these approaches are not used by any security-critical library that we investigated. Nevertheless, our analysis could be extended to binaries produced by these approaches.

\paragraph{Related Studies}
Compiler-induced secret-dependent operations have been studied in the past. They have been reproduced using toy examples~\cite{simon2018you,daniel2020binsec}. The toy examples mostly focus on binary arithmetic and how compilers may simplify these. Binsec/Rel~\cite{daniel2020binsec} also performs a study of several constant-time cryptographic implementations on the binary level on x86 and arm architectures. However, they do not investigate other architectures nor how other compilers can affect these implementations. To the best of our knowledge, we are the first to demonstrate compiler-induced issues in constant-time cryptographic libraries in the wild.

Kaufmann et al. discovered a secret dependency in a curve25519 implementation~\cite{kaufmann2016constant}. They detect that the Windows runtime library introduces a secret-dependent variation using the compiler MSVC 2015. In this paper, we do not trace the runtime library and will not flag such issues. All of the secret dependencies that we found purely originate from the library itself. We remark that issues in underlying runtime libraries are no less critical, and our approach could also be extended to also trace them.

\section{Conclusion}

General-purpose compilers can interfere with hardening techniques used to write high-level constant-time code, e.g., in cryptographic libraries.
In this paper, we investigated whether this happens in the wild and to what extent.
For this, we developed a flexible pipeline based on trace-based dynamic analysis that allowed us to measure a large number of target libraries, compilers, optimization versions, and processor architectures.
Using our pipeline, we analyzed whether the resulting binaries and algorithms exhibited any of the following two behaviors: control-flow decisions and memory accesses based on secret values, both of which are symptomatic of non-constant-time execution.

Our study demonstrates that compiler-induced side-channel vulnerabilities are widespread in cryptographic primitives of libraries that adopt constant-time programming techniques.
We observed issues in all tested processor architectures, compiler versions, and optimization levels.
These issues emerged in 5 of the 8 profiled libraries -- even one where the source code was formally verified to be free of such side channels.
We analyze a few notable cases in detail: the main secret-dependent operations introduced by the compilers are arithmetic shortcuts to skip some computation (e.g., early exit from loops), replacing bitmask arithmetic with branches, and splitting complex branching conditions -- directly reversing the recommendations of defensive programming guidelines to achieve constant-time code~\cite{aumasson2019cryptocoding}.
This suggests that compilers silently ``undo'' the code patterns that developers introduce purposefully to ensure constant-time execution.

Current principled defenses do not seem ready for mass adoption: compilers that provide constant-time code guarantees are specialized and lag behind general-purpose ones in terms of supported architectures, features, and performance. Formally verifying the assembly produced by compilers requires high manual efforts and is thus reserved for the most popular architectures only -- confirmed by our findings that less popular architectures are more subjected to compiler-induced vulnerabilities. Distributing certified binaries would greatly increase the efforts of developers and users of libraries alike.

Our work highlights that current defensive programming techniques are more akin to anecdotal workarounds to compilers' behavior rather than a systematic solution. Our results point out that high-level constant-time code is prone to dangerous transformations and optimizations from compilers and calls for cooperation between developers of security-critical libraries and general-purpose compilers.

\begin{acks}
    This work was supported by the \grantsponsor{zisc}{Zurich Information Security and Privacy Center}{https://zisc.ethz.ch/}, and \grantsponsor{efcl}{ETH Future Computing Laboratory}{https://efcl.ethz.ch/} (financed by a donation from Huawei Technologies).
\end{acks}

\citestyle{acmnumeric}

\bibliographystyle{abbrv}
\bibliography{bibliography}

\appendix
\section{Appendix}

\label{app:disclosure}

In this appendix, we state the data availability of the tools and data we generated in this paper. We then list the initial disclosure statements to the library developers in verbatim (barring some anonymization for the submission). We note that during the disclosure process, the developers sometimes asked for more tests on the patch, different versions, or on even more implementations of theirs. We complied with all of these requests and added these results to this paper. For instance, the HACL* developers told us about their RSA implementation, which we subsequently tested and added to this paper. 

\subsection{Data Availability}
We will release all research artifacts after publication under an open-source license. This includes the tool we developed to analyze binaries and the software infrastructure to compile a wide variety of crypto libraries with various compilers.

\subsection{HACL* Disclosure}

We ran a large-scale study of compiler-induced secret-dependent
operations on HACL* using dynamic analysis. The study covers 6 processor
architectures (x86-64, x86-i386, armv7, aarch64, mips32, RISC-V), two
compilers (LLVM and GCC), 17 compiler versions, and 7
optimization levels. HACL* was one of the targets of our study. 

Our dynamic analysis framework calls the API of HACL* with different
keys and fixes all other randomness. Then we collected a set of traces
which we then compared. Any difference in the traces is highlighted as a
potential secret-dependent operation. We did not find any issue in most
of our experiments of HACL*.

However, we found a code pattern in the secp256 implementation of HACL*
that is optimized by some compilers. The resulting
binary then contains secret-dependent operations, potentially leading to
a side-channel vulnerability.

We understand that you advise people to use formally verified compilers
such as CompCert. However, many people use your implementations with
off-the-shelf compilers. Therefore, it would be nice if this could be
fixed.

We tested the older version of the secp256 implementation
(\texttt{f283af1}), and
the following analysis and examples are taken from there. We are
currently rerunning our analysis for the current master (\texttt{3e283ef}), and
we have already found some preliminary secret-dependent operations in
the new version. We believe it stems from the same reason as the
original code snippet and thus first discuss the old one here.

\hypertarget{cmov}{%
\subsubsection{CMOV}\label{cmov}}

The \texttt{cmovznz4} function is supposed to be
independent of the input condition \texttt{cin}.
However, LLVM versions 13, 14, and 15 compile the snippet to
secret-dependent branches (on RISC-V) and secret-dependent memory
operations on MIPS, x86-i386, and armv7. An example of this can be seen
here: \url{https://godbolt.org/z/z71Wv5b7j}. A table showing the affected compiler versions, architectures, and optimization levels can be seen in \cref{tab:app:hacl:cmov}.

\begin{table}[tbp]
    \centering
    \begin{tabular}{@{}llllllll@{}}
    \toprule
    arch & toolchain & -O1 & -O2 & -O3 & -Ofast & -Os & -Oz\tabularnewline
    \midrule
    armv7 & LLVM 14 & \no & \no & \no & \no & \yes & \yes\tabularnewline
    & LLVM 15 & \no & \no & \no & \no & \yes & \yes\tabularnewline
    mips32el & LLVM 13 & \no & \no & \no & \no & \yes & \yes\tabularnewline
    & LLVM 14 & \no & \no & \no & \no & \yes & \yes\tabularnewline
    & LLVM 15 & \no & \no & \no & \no & \yes & \yes\tabularnewline
    riscv64 & LLVM 13 & \yes & \yes & \yes & \yes & \yes & \yes\tabularnewline
    & LLVM 14 & \yes & \yes & \yes & \yes & \yes & \yes\tabularnewline
    & LLVM 15 & \no & \no & \no & \no & \yes & \yes\tabularnewline
    x86-i686 & LLVM 14 & \no & \no & \no & \no & \yes & \yes\tabularnewline
    & LLVM 15 & \no & \no & \no & \no & \yes & \yes\tabularnewline
    \bottomrule
    \end{tabular}
    \caption{cmovnz in HACL*}
    \label{tab:app:hacl:cmov}
\end{table}

\hypertarget{current-master}{%
\subsubsection{Current Master}\label{current-master}}

A similar code pattern to the one shown above is also used in the function \texttt{point\_mul} in \\
\texttt{dist/gcc-compatible/Hacl\_P256.c}. Our analysis
also indicates that the code path is being executed with some dependency
on the secret key. However, this function is vastly more complex and
thus harder to pinpoint where the issue comes from. We believe it stems
from the same code pattern as above, but we are not 100\% confident.

A minimal reproducible example can be found here:
\url{https://godbolt.org/z/6s9daKhxr}. The secret-dependent branching
instruction that our analysis finds is
\texttt{jne .LBB0\_28} in line 609. A table showing the affected compiler versions, architectures, and optimization levels can be seen in \cref{tab:app:hacl:master}.

\begin{table}[tbp]
    \centering
    \begin{tabular}{@{}llllllll@{}}
    \toprule
    arch & toolchain & -O1 & -O2 & -O3 & -Ofast & -Os & -Oz\tabularnewline
    \midrule
    aarch64 & LLVM 13 & \yes & \no & \no & \no & \no & \no\tabularnewline
    & LLVM 14 & \yes & \no & \no & \no & \no & \no\tabularnewline
    & LLVM 15 & \yes & \no & \no & \no & \no & \no\tabularnewline
    armv7 & LLVM 13 & \yes & \no & \no & \no & \no & \no\tabularnewline
    & LLVM 14 & \yes & \yes & \yes & \yes & \yes & \yes\tabularnewline
    & LLVM 15 & \yes & \yes & \yes & \yes & \yes & \yes\tabularnewline
    mips32el & LLVM 13 & \yes & \yes & \yes & \yes & \yes & \yes\tabularnewline
    & LLVM 14 & \yes & \yes & \yes & \yes & \yes & \yes\tabularnewline
    & LLVM 15 & \yes & \yes & \yes & \yes & \yes & \yes\tabularnewline
    riscv64 & LLVM 13 & \yes & \yes & \yes & \yes & \yes & \yes\tabularnewline
    & LLVM 14 & \yes & \yes & \yes & \yes & \yes & \yes\tabularnewline
    & LLVM 15 & \yes & \yes & \yes & \yes & \no & \no\tabularnewline
    x86-64 & LLVM 13 & \yes & \yes & \yes & \yes & \yes & \yes\tabularnewline
    & LLVM 14 & \yes & \yes & \yes & \yes & \yes & \yes\tabularnewline
    & LLVM 15 & \yes & \yes & \yes & \yes & \no & \no\tabularnewline
    x86-i686 & LLVM 13 & \yes & \no & \no & \no & \no & \no\tabularnewline
    & LLVM 14 & \yes & \yes & \yes & \yes & \yes & \yes\tabularnewline
    & LLVM 15 & \yes & \yes & \yes & \yes & \yes & \yes\tabularnewline
    \bottomrule
    \end{tabular}
    \caption{cmovnz in HACL* (current master)}
    \label{tab:app:hacl:master}
\end{table}

\subsection{BoringSSL Disclosure}

We ran a large-scale study of compiler-induced secret-dependent
operations on multiple cryptographic libraries using dynamic analysis.
The study covers 6 processor architectures (x86-64, x86-i386, armv7,
aarch64, mips32, RISC-V), two compilers (LLVM and GCC), a total of 17
compiler versions, and 7 optimization levels. BoringSSL was one of the
targets of our study. 
We only
investigate secret-dependent memory accesses and secret-dependent
control flow.

Our dynamic analysis framework uses emulation to execute binaries for
different target architectures. We call some of the APIs exposed by
BoringSSL in the binaries. To find secret-dependent operations, our tool
executes the binaries with different secret keys but fixes all other
sources of randomness. We then collect a trace of the executed binaries
for each secret key. Any difference in the traces is highlighted as a
potential secret-dependent operation. Such secret-dependent operations
can lead to an exploitable side-channel vulnerability. However, we do
not know if the secret-dependent operations that we detect are
exploitable at all. Nevertheless, you might be interested in them.

We used the branch \texttt{chromium-stable} or
28f96c26 for our analysis.

In general, we did not find any issue that was present in all
architectures and all compilers. Such results would indicate issues at
the source-code level. However, we found several secret-dependent
operations that seem compiler induced as they only appear for some
combination of compiler, architecture, and optimization levels. The
resulting binary then contains secret-dependent operations, potentially
leading to a side-channel vulnerability.

\hypertarget{bn_mul_comba8-and-bn_sqr_comba8}{%
\subsubsection{\texorpdfstring{\texttt{bn\_mul\_comba8} and
\texttt{bn\_sqr\_comba8}}{bn\_mul\_comba8 and bn\_sqr\_comba8}}\label{bn_mul_comba8-and-bn_sqr_comba8}}

Our analysis found secret-dependent operations in the function
\texttt{bn\_mul\_comba8} and
\texttt{bn\_sqr\_comba8} on MIPS (mips32el). Both
functions are executed during ECDH-P256 and ECDSA. To reproduce it and
to demonstrate the secret-dependent operation, we created a small
reproducible example using \texttt{bn\_mul\_comba8}:
\url{https://godbolt.org/z/1of34165o}. The resulting binary on the right
contains multiple branching instructions that depend on the parameters
of this function. One such branch can be found on line 52. Based on the
source code, we assume the issue would also pop up in other 32-bit
architectures. However, we did not observe that with x86-i386 and armv7. A table showing the affected compiler versions, architectures, and optimization levels can be seen in \cref{tab:app:boringssl:bnmul}.

\begin{table}[tbp][tp]
    \centering
    \begin{tabular}{@{}lllllllll@{}}
    \toprule
    Arch & Compiler & O0 & O1 & O2 & O3 & Ofast & Os & Oz\tabularnewline
    \midrule
    mips32el & GCC 5.4.0 & \yes & \no & \yes & \yes & \yes & \no &
    \no\tabularnewline
     & GCC 6.3.0 & \yes & \no & \yes & \yes & \yes & \no &
    \no\tabularnewline
     & GCC 7.3.0 & \yes & \no & \yes & \yes & \yes & \no &
    \no\tabularnewline
     & GCC 8.4.0 & \yes & \no & \yes & \yes & \yes & \no &
    \no\tabularnewline
     & GCC 9.3.0 & \yes & \no & \yes & \yes & \yes & \no &
    \no\tabularnewline
     & GCC 10.3.0 & \yes & \no & \yes & \yes & \yes & \no &
    \no\tabularnewline
     & GCC 11.3.0 & \yes & \no & \yes & \yes & \yes & \no &
    \no\tabularnewline
     & LLVM 11 & \yes & \no & \no & \no & \no & \no & \no\tabularnewline
     & LLVM 13 & \yes & \no & \no & \no & \no & \no & \no\tabularnewline
     & LLVM 14 & \yes & \no & \no & \no & \no & \no & \no\tabularnewline
     & LLVM 15 & \yes & \no & \no & \no & \no & \no & \no\tabularnewline
    \bottomrule
    \end{tabular}
    \caption{\texttt{bn\_mul\_comba8} and \texttt{bn\_sqr\_comba8} in BoringSSL}
    \label{tab:app:boringssl:bnmul}
\end{table}

\hypertarget{bn_sqr_comba4}{%
\subsubsection{\texorpdfstring{\texttt{bn\_sqr\_comba4}}{bn\_sqr\_comba4}}\label{bn_sqr_comba4}}

We found a similar issue when compiling BoringSSL on RISC-V (riscv64) in
the function \texttt{bn\_sqr\_comba4}. A small
reproducible example can be found in
\url{https://godbolt.org/z/EGdnWhzsz}. Note the branching
instruction on line 18.

We observed this behavior with the combinations of
compilers, architectures, and optimization levels listed in \cref{tab:app:boringssl:bnsqr}.

\begin{table}[tp]
    \centering
    \begin{tabular}{@{}lllllllll@{}}
    \toprule
    Arch & Compiler & O0 & O1 & O2 & O3 & Ofast & Os & Oz\tabularnewline
    \midrule
    riscv64 & GCC 10.3.0 & \yes & \no & \yes & \yes & \yes & \no &
    \no\tabularnewline
     & GCC 11.3.0 & \yes & \no & \yes & \yes & \yes & \no &
    \no\tabularnewline
    \bottomrule
    \end{tabular}
    \caption{\texttt{bn\_sqr\_comba4} in BoringSSL}
    \label{tab:app:boringssl:bnsqr}
\end{table}

\hypertarget{fiat_p256_point_add}{%
\subsubsection{\texorpdfstring{\texttt{fiat\_p256\_point\_add}}{fiat\_p256\_point\_add}}\label{fiat_p256_point_add}}

Our analysis flagged a branch in
\texttt{fiat\_p256\_point\_add} in 32bit architectures
(i.e., armv7, x86-i386, and mips32el). However, this function is
significantly more complex, which makes it hard to pinpoint the problem.
It seems as if the if condition
\texttt{if \string(is\_nontrivial\_double\string)} gets compiled into
two distinct branching instructions instead of a single one, and which
of these is taken depends on the secret key.

Here is a ``short'' reproduction of the function:
\url{https://godbolt.org/z/sTGKdbEcK}. We believe LLVM inserts two distinct
branches instead of one (line 116 and 120 respectively), and which of
these two is taken leaks some information on the input.

\Cref{tab:app:boringssl:fiat} lists the combinations of
compilers, architectures, and optimization levels with which we observed this behavior.

\begin{table}[tp]
    \centering
    \begin{tabular}{@{}lllllllll@{}}
    \toprule
    Arch & Compiler & O0 & O1 & O2 & O3 & Ofast & Os & Oz\tabularnewline
    \midrule
    armv7 & LLVM 14 & \no & \yes & \yes & \yes & \yes & \yes & \yes\tabularnewline
     & LLVM 15 & \no & \yes & \yes & \yes & \yes & \yes & \yes\tabularnewline
    mips32el & LLVM 14 & \no & \yes & \yes & \yes & \yes & \yes &
    \yes\tabularnewline
     & LLVM 15 & \no & \yes & \yes & \yes & \yes & \yes &
    \yes\tabularnewline
    x86-i386 & LLVM 14 & \no & \yes & \yes & \yes & \yes & \yes &
    \yes\tabularnewline
     & LLVM 15 & \no & \yes & \yes & \yes & \yes & \yes &
    \yes\tabularnewline
    \bottomrule
    \end{tabular}
    \caption{\texttt{fiat\_p256\_point\_add} in BoringSSL}
    \label{tab:app:boringssl:fiat}
\end{table}

\subsection{Botan Disclosure}

We ran a large-scale study of compiler-induced secret-dependent
operations on multiple cryptographic libraries using dynamic analysis.
The study covers 6 processor architectures (x86-64, x86-i386, armv7,
aarch64, mips32, RISC-V), two compilers (LLVM and GCC), a total of 17
compiler versions, and 7 optimization levels. Botan was one of the
targets of our study. 
We only
investigate secret-dependent memory accesses and secret-dependent
control flow.

Our dynamic analysis framework uses emulation to execute binaries for
different target architectures. We call some of the APIs exposed by
Botan in the binaries. To find secret-dependent operations, our tool
executes the binaries with different secret keys but fixes all other
sources of randomness. We then collect a trace of the executed binaries
for each secret key. Any difference in the traces is highlighted as a
potential secret-dependent operation. Such secret-dependent operations
can lead to an exploitable side-channel vulnerability. However, we do
not know if the secret-dependent operations that we detect are
exploitable at all. Nevertheless, you might be interested in them.

We analyzed the release 2.19.3 (\texttt{15dc32f}).

In general, we did not find any issue that was present in all
architectures and all compilers. Such results would indicate issues at
the source-code level. However, we found several secret-dependent
operations that seem compiler induced as they only appear for some
combination of compiler, architecture, and optimization levels. The
resulting binary then contains secret-dependent operations, potentially
leading to a side-channel vulnerability.

\hypertarget{hex_decode}{%
\subsubsection{\texorpdfstring{\texttt{hex\_decode}}{hex\_decode}}\label{hex_decode}}

Our analysis found secret-dependent operations in the function
\texttt{hex\_decode} on recent LLVM versions targeting
riscv64 and x86-i386. This function is flagged in various cryptographic
algorithms: multiple modes of AES, chachapoly1305, curve25519, and HMAC
with multiple hash functions. To reproduce it and to demonstrate the
secret-dependent operation, we created a small reproducible example:
\url{https://godbolt.org/z/hhs1KWfz7}. The resulting binary on the right
contains multiple branching instructions that depend on the parameters
of this function. One such branch can be found on line 54.

\Cref{tab:app:botan:hex} lists the combinations of
compilers, architectures, and optimization levels with which we observed this behavior.

\begin{table}[tp]
    \centering
    \begin{tabular}{@{}lllllllll@{}}
    \toprule
    Arch & Compiler & O0 & O1 & O2 & O3 & Ofast & Os & Oz\tabularnewline
    \midrule
    riscv64 & LLVM 14 & \no & \yes & \yes & \yes & \yes & \yes &
    \yes\tabularnewline
     & LLVM 15 & \no & \yes & \yes & \yes & \yes & \yes &
    \yes\tabularnewline
    x86-i386 & LLVM 14 & \no & \yes & \yes & \yes & \yes & \yes &
    \yes\tabularnewline
     & LLVM 15 & \no & \yes & \yes & \yes & \yes & \yes &
    \yes\tabularnewline
    \bottomrule
    \end{tabular}
    \caption{\texttt{hex\_decode} in Botan}
    \label{tab:app:botan:hex}
\end{table}

\hypertarget{custom-128bit-type}{%
\subsubsection{Custom 128bit type}\label{custom-128bit-type}}

Our analysis flaggs a custom 128bit type and one of its operator as
secret-dependent in the implementation of poly1305 and curve25519. More
specifically, the \texttt{donna128} and its
\texttt{+=} operator is by some compilers optimized
into secret-dependent operations. We replicated the issue and created a
minimal reproducible example: \url{https://godbolt.org/z/P3n3vWsdT}. Note that
some compilers only sometimes introduce the branch to skip the carry
addition. E.g., in x86-i386, we observed that the compiler did not always
produce the same inline binary for the addition operator. For example,
\url{https://godbolt.org/z/xv9fznMfK} is a larger example using x86-i386 in
the function \texttt{fsquare} where the compiler
inserts the branch in line 193.

\Cref{tab:app:botan:donna} lists the combinations of
compilers, architectures, and optimization levels with which we observed this behavior.

\begin{table}[tp]
    \centering
    \begin{tabular}{@{}lllllllll@{}}
    \toprule
    Arch & Compiler & O0 & O1 & O2 & O3 & Ofast & Os & Oz\tabularnewline
    \midrule
    mips32el & GCC 6.4.0 & \yes & \yes & \yes & \yes & \yes & \yes &
    \no\tabularnewline
     & GCC 7.3.0 & \yes & \yes & \yes & \yes & \yes & \yes &
    \no\tabularnewline
     & GCC 8.4.0 & \yes & \yes & \yes & \yes & \yes & \yes &
    \no\tabularnewline
     & GCC 9.3.0 & \yes & \yes & \yes & \yes & \yes & \yes &
    \no\tabularnewline
     & GCC 10.3.0 & \yes & \yes & \yes & \yes & \yes & \yes &
    \no\tabularnewline
     & GCC 11.3.0 & \yes & \yes & \yes & \yes & \yes & \yes &
    \no\tabularnewline
    x86-i386 & GCC 7.3.0 & \yes & \yes & \yes & \yes & \yes & \yes &
    \no\tabularnewline
     & GCC 8.4.0 & \yes & \yes & \yes & \yes & \yes & \no &
    \no\tabularnewline
     & GCC 9.3.0 & \yes & \yes & \no & \no & \no & \no & \no\tabularnewline
     & GCC 10.3.0 & \yes & \yes & \no & \no & \no & \no & \no\tabularnewline
     & GCC 11.3.0 & \yes & \yes & \no & \yes & \yes & \no &
    \no\tabularnewline
    \bottomrule
    \end{tabular}
    \caption{Custom 128-bit type in Botan}
    \label{tab:app:botan:donna}
\end{table}

\subsubsection{\texttt{key\_schadule} in GHASH}
\label{key_schadule-in-ghash}

Our analysis flagged another secret-dependent operation in the function
\texttt{key\_schedule} in GHASH. A reproducible example
can be found here: \url{https://godbolt.org/z/xGh6Kjzfh}. Essentially,
the compiler understands that the \texttt{carry}
variable can only be 1 or 0 and thus checks its value before it performs
the XOR.

We found this behavior with LLVM 15 for riscv64 with
\texttt{-O1}, \texttt{-O2},
\texttt{-O3}, and \texttt{-Ofast}.

\subsection{BearSSL Disclosure}

We ran a large-scale study of compiler-induced secret-dependent
operations on multiple cryptographic libraries using dynamic analysis.
The study covers 6 processor architectures (x86-64, x86-i386, armv7,
aarch64, mips32, RISC-V), two compilers (LLVM and GCC), a total of 17
compiler versions, and 7 optimization levels. BearSSL was one of the
targets of our study. 
We only
investigate secret-dependent memory accesses and secret-dependent
control flow.

Our dynamic analysis framework uses emulation to execute binaries for
different target architectures. We call some of the APIs exposed by
BearSSL in the binaries. To find secret-dependent operations, our tool
executes the binaries with different secret keys but fixes all other
sources of randomness. We then collect a trace of the executed binaries
for each secret key. Any difference in the traces is highlighted as a
potential secret-dependent operation. Such secret-dependent operations
can lead to an exploitable side-channel vulnerability. However, we do
not know if the secret-dependent operations that we detect are
exploitable at all. Nevertheless, you might be interested in them.

We used the branch \texttt{master} on
79c060ee for our analysis.

In general, we found one issue that was present in all architectures
and all compilers. This indicates that the issue is at the source-code
level. In addition, we found several secret-dependent operations that
seem compiler induced as they only appear for some combination of
compiler, architecture, and optimization level. The resulting binary
then contains secret-dependent operations, potentially leading to a
side-channel vulnerability.

\hypertarget{br_skey_decoder_run}{%
\subsubsection{\texorpdfstring{\texttt{br\_skey\_decoder\_run}}{br\_skey\_decoder\_run}}\label{br_skey_decoder_run}}

We load the secret keys from a file using
\texttt{br\_skey\_decoder\_run}, which seems to contain
operations that depend on the content of the file. We made sure that our
inputs were all the same type of key (i.e., RSA) with different content.
We found these issues in every binary, so we assume they are rooted in
source-code and not introduced by compiler optimizations. We also did
not further evaluate where these issues come from, but you might be
interested in them either way.

\hypertarget{modpow_opt}{%
\subsubsection{\texorpdfstring{\texttt{modpow\_opt}}{modpow\_opt}}\label{modpow_opt}}

The function \texttt{modpow\_opt} in both \texttt{i31} and \texttt{i62} were flagged to contain
compiler-induced issues on some architectures and compilers. We created
small snippets that demonstrate the issue for
\texttt{i31} and \texttt{i62}
respectively:

\begin{itemize}
\item
  \texttt{i31}: \url{https://godbolt.org/z/bnE1PrfKG}
\item
  \texttt{i62}: \url{https://godbolt.org/z/nx1crrxv3}
\end{itemize}

Essentially, the compiler understands the bitmasks computation that is
performed using \texttt{EQ} and then
optimizes/simplifies the constant-time assignment. In
\texttt{i31} the compiler uses a branch to jump over
the assignment and in \texttt{i62} the compiler
computes the address that is then used for the assignment (the address
is then leaked in the actual memory access, i.e., line 10 and 11 in the
assembly output of the Godbolt link above).

\Cref{tab:app:bearssl:modpow} lists the combinations of
compilers, architectures, and optimization levels with which we observed this behavior.

\begin{table}[tp]
    \centering
    \begin{tabular}{@{}lllllllll@{}}
    \toprule
    Arch & Compiler & O0 & O1 & O2 & O3 & Ofast & Os & Oz\tabularnewline
    \midrule
    aarch64 & LLVM 13 & \no & \no & \yes & \no & \no & \no & \no\tabularnewline
     & LLVM 14 & \no & \no & \yes & \no & \no & \yes & \yes\tabularnewline
     & LLVM 14 & \no & \yes & \no & \no & \no & \yes & \yes\tabularnewline
    armv7 & LLVM 13 & \no & \no & \yes & \yes & \yes & \no & \no\tabularnewline
    x86-64 & LLVM 13 & \no & \yes & \yes & \yes & \yes & \yes & \yes\tabularnewline
     & LLVM 14 & \no & \yes & \yes & \yes & \yes & \yes & \yes\tabularnewline
     & LLVM 15 & \no & \yes & \yes & \yes & \yes & \yes & \yes\tabularnewline
    x86-i386 & LLVM 13 & \no & \yes & \yes & \yes & \yes & \yes &
    \yes\tabularnewline
     & LLVM 14 & \no & \yes & \yes & \yes & \yes & \yes &
    \yes\tabularnewline
     & LLVM 15 & \no & \yes & \yes & \yes & \yes & \yes &
    \yes\tabularnewline
    \bottomrule
    \end{tabular}
    \caption{\texttt{modpow\_opt} in BearSSL}
    \label{tab:app:bearssl:modpow}
\end{table}

\hypertarget{br_i31_muladd_small}{%
\subsubsection{\texorpdfstring{\texttt{br\_i31\_muladd\_small}}{br\_i31\_muladd\_small}}\label{br_i31_muladd_small}}

The compiler once again seems to understand the bitwise arithmetic
operations of \texttt{MUX},
\texttt{EQ}, and \texttt{GT} and
produces secret-dependent operations in the function \\
\texttt{br\_i31\_muladd\_small}. A small reproducible
code snippet can be found here: \url{https://godbolt.org/z/qM9KT31vz}.

\Cref{tab:app:bearssl:muladd} lists the combinations of
compilers, architectures, and optimization levels with which we observed this behavior.

\begin{table}[tp]
    \centering
    \begin{tabular}{@{}lllllllll@{}}
    \toprule
    Arch & Compiler & O0 & O1 & O2 & O3 & Ofast & Os & Oz\tabularnewline
    \midrule
    x86-i386 & LLVM 13 & \no & \yes & \yes & \yes & \yes & \no & \no\tabularnewline
     & LLVM 14 & \no & \yes & \yes & \yes & \yes & \no &
    \yes\tabularnewline
     & LLVM 15 & \no & \yes & \yes & \yes & \yes & \no & \no\tabularnewline
    riscv64 & LLVM 13 & \no & \yes & \yes & \yes & \yes & \no & \no\tabularnewline
     & LLVM 14 & \no & \yes & \yes & \yes & \yes & \no & \yes\tabularnewline
     & LLVM 15 & \no & \yes & \yes & \yes & \yes & \no & \no\tabularnewline
    \bottomrule
    \end{tabular}
    \caption{\texttt{br\_i31\_muladd\_small} in BearSSL}
    \label{tab:app:bearssl:muladd}
\end{table}

\hypertarget{p256-check_scalar}{%
\subsubsection{\texorpdfstring{P256
\texttt{check\_scalar}}{P256 check\_scalar}}\label{p256-check_scalar}}

We found a misguided compiler optimization in the function
\texttt{check\_scalar} in the p256 elliptic curve
implementation (both for \texttt{i62} and
\texttt{i64}). A small reproducible example can be
found here: \url{https://godbolt.org/z/zvh6Pjzar}.

\Cref{tab:app:bearssl:check} lists the combinations of
compilers, architectures, and optimization levels with which we observed this behavior.

\begin{table}[tp]
    \centering
    \begin{tabular}{@{}lllllllll@{}}
    \toprule
    Arch & Compiler & O0 & O1 & O2 & O3 & Ofast & Os & Oz\tabularnewline
    \midrule
    x86-i386 & LLVM 12 & \no & \no & \no & \no & \no & \yes & \yes\tabularnewline
     & LLVM 13 & \no & \yes & \yes & \yes & \yes & \yes &
    \yes\tabularnewline
     & LLVM 14 & \no & \yes & \yes & \yes & \yes & \yes &
    \yes\tabularnewline
     & LLVM 15 & \no & \yes & \yes & \yes & \yes & \yes &
    \yes\tabularnewline
    riscv64 & LLVM 12 & \no & \no & \yes & \yes & \yes & \no & \no\tabularnewline
     & LLVM 13 & \no & \yes & \yes & \yes & \yes & \no & \no\tabularnewline
     & LLVM 14 & \no & \yes & \yes & \yes & \yes & \no & \yes\tabularnewline
     & LLVM 15 & \no & \yes & \yes & \yes & \yes & \no & \no\tabularnewline
    \bottomrule
    \end{tabular}
    \caption{\texttt{check\_scalar} in BearSSL}
    \label{tab:app:bearssl:check}
\end{table}

\newpage

\end{document}